\documentclass[conferences]{IEEEtran}
\usepackage{array}
\usepackage{graphicx}
\usepackage{amsmath}
\usepackage{algorithmic}
\usepackage{algorithm}
\usepackage{url}
\usepackage{tikz}
\usepackage{cite}
\usepackage[font={small},skip=5pt]{caption}
\usepackage{subcaption}
\usepackage{setspace}
\usepackage{enumitem}
\usepackage{graphicx,lipsum}
\usepackage{caption}

\renewcommand{\arraystretch}{1.0}  
\newcolumntype{P}[1]{>{\centering\arraybackslash}p{#1}} 
\newcolumntype{M}[1]{>{\centering\arraybackslash}m{#1}} 

\setenumerate[1]{label=(\alph*)}
\setenumerate[0]{label=(\roman*)}
\def\thesection{\arabic{section}}
\newcommand\copyrighttext{%
  \footnotesize \centering \copyright~2015 Springer. The final publication is available at Springer via http://dx.doi.org/10.1007/s11276-015-1002-4.}
\newcommand\copyrightnotice{%
\begin{tikzpicture}[remember picture,overlay]
\node[anchor=south,yshift=10pt] at (current page.south) {\fbox{\parbox{\dimexpr\textwidth-\fboxsep-\fboxrule\relax}{\copyrighttext}}};
\end{tikzpicture}%
}

\begin{document}
\bibliographystyle{IEEEtran}
%

\title{Interference Mitigation In Wireless Mesh Networks Through Radio Co-location Aware Conflict Graphs}

\author{\IEEEauthorblockN{Srikant Manas Kala\IEEEauthorrefmark{1}, M Pavan Kumar Reddy, Ranadheer Musham, Bheemarjuna Reddy Tamma\\}
 \IEEEauthorblockA{ Indian Institute of Technology Hyderabad, India\\
 Email:[ \IEEEauthorrefmark{1}cs12m1012, cs12b1025, cs12b1026, tbr ]@iith.ac.in}}
%


\maketitle

\begin{abstract}
Wireless Mesh Networks (WMNs) have evolved into a wireless communication technology of immense interest. But technological advancements in WMNs have inadvertently spawned a plethora of network performance bottlenecks, caused primarily by the rise in prevalent interference. Conflict Graphs are indispensable tools used to theoretically represent and estimate the interference in wireless networks. We propose a generic algorithm to generate conflict graphs which is independent of the underlying interference model. Further, we propose the notion of radio co-location interference, which is caused and experienced by spatially co-located radios in multi-radio multi-channel (MRMC) WMNs. We experimentally validate the concept, and propose a new all-encompassing algorithm to create a radio co-location aware conflict graph. Our novel conflict graph generation algorithm is demonstrated to be significantly superior and more efficient than the conventional approach, through theoretical interference estimates and 
comprehensive experiments. The results of an extensive set of ns-3 simulations run on the IEEE 802.11g platform strongly indicate that the radio co-location aware conflict graphs are a marked improvement over their conventional counterparts. We also question the use of \textit{total interference degree} as a reliable metric to predict the performance of a Channel Assignment scheme in a given WMN deployment.
\end{abstract}




\section{Introduction}
\begin{figure}[htb!]
                \centering
                \includegraphics[width=8.5cm, height=7cm]{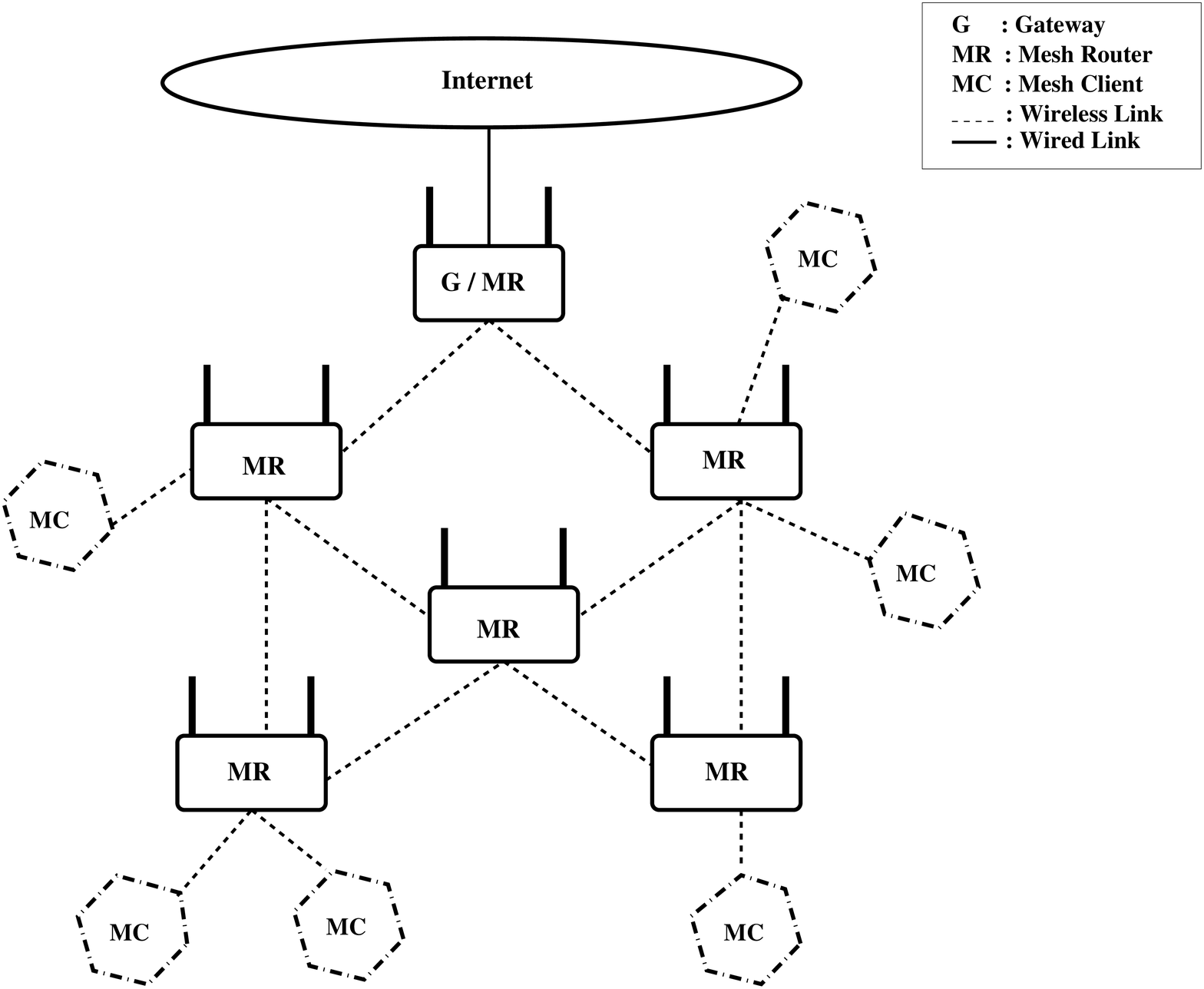}
                \caption{A Simplistic WMN Architecture}
                \label{WMN}
        \end{figure}

Wireless Mesh Networks have emerged as a promising technology, with a potential for widespread application in contemporary wireless networks. They have the potential to substitute, and thereby reduce the dependence on the wired infrastructure. In the foreseeable future, WMNs may be extensively deployed due to consistently increasing low-cost availability of the commodity IEEE 802.11 off-the-shelf hardware, smooth deployment with ease of scalability, effortless reconfigurability and increased network coverage \cite{10Wang}\cite{11Bruno}. The surge in their presence will be equally attributed to the tremendous increase in data communication rates that are being guaranteed by the IEEE 802.11 and IEEE 802.16 standards. WMNs also offer enhanced reliability when compared to their wired counterparts because of the inherent redundancy in the underlying mesh topology. 
WMN technology, given its practical and commercial appeal, can adequately cater to the needs of a variety of network applications. These networks range from institutional and social wireless LANs, last-mile broadband Internet access, to disaster networks. Prominent wireless technologies that stand to benefit from WMN deployments, other than the IEEE 802.11 WLANs, are the IEEE 802.16 Wireless Metropolitan Area Networks (WMANs) and the next generation cellular mobile systems, including LTE-Advanced \cite{12Capone}. WMNs are also poised to form the backbone of the next-generation of integrated wireless networks, that aim to converge a plethora of technologies such as 3G/4G mobile networks, WLANs etc. onto a single communication delivery platform \cite{15Skalli}.
\copyrightnotice
The mesh topology framework in a WMN facilitates multiple-hop transmissions to relay the data traffic seamlessly between source-destination pairs that are often beyond the transmission range of each other \cite{6Akyildiz}. Thus each node in the \textit{Wireless Mesh Backbone} acts as a host and as a router, forwarding packets onto the next hop. A WMN deployment can provide both, a self-contained IEEE 802.11 WLAN with no connectivity to foreign networks, as well as an unrestricted access to outside networks, through a Gateway. Several Gateways may be required if the WMN has to establish communication links with external networks. A simplistic WMN architecture is constituted of numerous mesh-routers which relay/route the data traffic via multiple-hop transmissions, and leverage the twin WMN features of being fully wireless and having a mesh topology. The mesh-clients are the ultimate end-user devices that are serviced by mesh-routers. The mesh-routers constitute the 
communication backbone of the WMN. Gateways exhibit operational duality by interfacing the WMN with outside networks, besides functioning as any other mesh-router within the WMN. IEEE 802.11 \cite{1IEEE} protocol standards serve as a popular link layer protocol for WMN deployment. A trivial single-gateway WMN is illustrated in Figure~\ref{WMN}, with mesh-routers and mesh-clients, which is the WMN model we adhere to in our research work. We consider the availability of multiple radios specifically for inter mesh-router communication, and do not deal with the mesh backbone to mesh-client communication issues. Hereafter, mesh-routers are referred to as “nodes”. 

Initial deployments of WMNs comprised of a trivial single-radio single-channel architecture, in which all nodes were equipped with a single radio and assigned the same channel. Subsequent performance analysis of such wireless network architectures revealed that there was substantial degradation in the network performance metrics, as the size of the WMN is scaled up \cite{2Gupta}. Single-channel deployments also adversely affect the end-to-end throughput and network capacity in IEEE 802.11 WMNs \cite{3Saadawi}. A slightly enhanced architecture is that of the single-radio multi-channel WMNs. But their performance in active deployments has been sub-par, due to the problems of disconnectivity in the WMN topology and the delays in channel switching. In a multi-channel network, assigning different channels to single-radio nodes leads to disruption of wireless connectivity between nodes, even though they lie within the transmission range of each other. Also, dynamically switching to the most suited 
channel based on the network performance indices, causes a delay of the order of milliseconds \cite{8Raniwala}. This switching delay is comparable to transmission delays and requires a near-perfect time synchronization between the nodes, so that they are on the same channel when they need to communicate \cite{15Skalli}.

The most efficient and pervasive WMN architecture is the MRMC framework. Availability of non-overlapping channels under the IEEE 802.11 and IEEE 802.16 standards, and cheap off-the-shelf wireless network interface cards, have propelled the deployment of MRMC  WMNs. The IEEE 802.11b/g/n standard utilizes the unlicensed $2.4$ GHz frequency band and provides $3$ orthogonal channels centered at $25$ MHz frequency spacing while the IEEE 802.11a/ac standard operates in the \textit{unlicensed national information infrastructure} band (U-NII band) that ranges from $5.15$ GHz to $5.85$ GHz \cite{14Weisheng}. The assured number of orthogonal channels range from $12$ to $24$, depending upon the channel bandwidth \emph{i.e.,} $20$ or $40$ MHz, and the country/region in context. Local laws and policies may restrict access to some frequencies of the $5$ GHz band and may also mandate the use of specific communication technologies in the restricted spectrum. For example, in the USA, devices 
operating in the $5$ GHz spectrum are permitted to use only a part of the spectrum, and are required to employ \textit{transmit power control} and \textit{dynamic frequency selection} techniques.
The presence of multiple radios on each node coupled with the availability of multiple channels, facilitates concurrent signal transmissions and receptions on the WMN nodes. Thus MRMC WMNs register an enhanced network capacity and an improved spectrum efficiency \cite{8Raniwala}\cite{9MeshDynamics}. However, the technological enhancements in WMN deployments have incidentally led to an undesired increase in the interference that impedes the radio communication in such wireless networks. 
\subsection{Research Problem Outline}
Numerous research studies have attempted to address and resolve interference related issues in WMNs. The phenomena of \textit{Radio co-location Interference} or RCI \emph{i.e.,} interference caused and experienced by spatially co-located radios or SCRs, that are operating on identical frequencies, is a crucial aspect of the multifaceted interference problem. But it has been largely unaddressed and finds very little mention in the current WMN research literature. We choose to focus primarily on this issue, and strive to adequately address and mitigate the adverse effects of RCI in a WMN. The first step towards mitigation of the detrimental affects of any interference bottleneck, is its correct identification and representation in the conflict graph. Thus, we accomplish our end goal by an accurate and wholesome representation of all possible RCI scenarios of a WMN in its \textit{conflict graph}.
\subsection{Paper Organization}
In Section $2$, we touch upon the important interference related themes such as categorization and representation of wireless interference and the \textit{interference models}. In Section $3$, we briefly mention some notable research literature relevant to our study. Section $4$ introduces the concept of RCI supplemented with two crucial RCI scenarios, and a \textit{proof of concept} experiment in support of the theoretical arguments. In Section $5$, we propose two generic multi-radio multi-channel conflict graph generation algorithms, \emph{viz.,} a conventional approach and a novel radio co-location aware approach. Section $6$ comprises of simulation methodologies, data traffic characteristics and presentation and analysis of recorded results. In Section $7$ we discuss the reliability of total interference degree as a theoretical estimate of prevalent interference. In Section $8$, we derive concrete conclusions accompanied with logical inferences based on the observed results and offered analysis. Finally, 
Section $9$ outlines the future course of our research work. 
  
\section{Interference In WMNs}
With the advent of MRMC deployments, the spectral complexity of the WMNs intensified. This led to a substantial rise in the interference endemic in WMNs, identification and mitigation of which continues to be the focus of researchers. 
\subsection{Categorizing Interference In A WMN}
Interference in a WMN can be broadly classified into the following three categories \cite{26Koutsonikolas}.
\begin{enumerate}
 \item\textbf{\textit{External}} : Interference caused by external wireless devices \emph{i.e.,} un-intentional interferers. It is uncontrolled as external wireless devices may communicate over a common frequency spectrum but are beyond the supervision of the MAC protocol employed in the WMN. Examples are Microwave ovens, Bluetooth devices, and other WMNs or WLANs operating in the same frequency band. 
 \item \textbf{\textit{Internal}} : Interference originating from within the WMN due to the broadcast nature of wireless communication. A transmission is generally isotropic and causes undesired interference at some of the neighboring nodes of the node for which it is intended. Network topology, channel allocation, and routing schemes have an immense impact on the intensity of internal interference.
 \item \textbf{\textit{Multipath Fading}} : It causes inter-symbol interference which occurs when the signals emanated from a particular source take multiple paths to arrive at the destination. The signals differ in time or phase at the destination and interfere with each other.
\end{enumerate}
In our work, we focus only on the internal or controlled interference, as it is the primary disruptive component of interference that leads to poor network performance in MRMC WMNs.

\subsection{Interference Models}
The next step would entail determining the conflicting wireless links in the WMN. This is a very complex problem due to the wireless nature of the network. Hence, researchers employ an accurate network interference model which is suitable for the WMN. The parameters of the model are used to ascertain the interfering or conflicting links. Selection of an appropriate model is crucial in representing complex wireless interference characteristics in a simple mathematical fashion. It is also pivotal in studying the impact on network performance and behavior, due to the adverse effects of interference.
The popular approaches taken to model the interference prevalent in wireless networks are elucidated below \cite{27Iyer}\cite{28Cardieri}.
\begin{enumerate}
 \item \textbf{\textit{The Physical or the Additive Interference Model}} : It is the closest representation of the actual physical interference experienced by radios, but is complex to formulate. It takes additive interference into account and considers a fixed \textit{signal to interference plus noise ratio} (SINR) threshold for successful data reception. Since it closely resembles a real-world interference scenario, it is a non-binary interference model. Thus a received signal may be attenuated due to interference, but as long as its strength exceeds the SINR threshold value, it is considered to be a successful transmission. 
 \item \textbf{\textit{The Capture Threshold Model}} : It is a simplified version of the Physical model which makes use of three threshold values instead of one. Further, the interference modeling is carried out separately for every interfering signal.
 \item \textbf{\textit{The Protocol Model}} : In this model, a transmission is successful if it does not experience any interference from other concurrent transmissions in its proximity. Every radio has a transmission and an interference range, where the latter is generally greater than the former. The interference range is usually 2-3 times the transmission range in actual deployments. The protocol model states that a signal transmission from radio $R_1$ to radio $R_2$ is deemed successful if $R_2$ falls within the transmission range of $R_1$, but not in the interference range of any other radio which may be active and transmitting concurrently.
Consider a graph $G =(V,E)$ which represents a WMN. Here V denotes the set of all nodes in the WMN and E denotes the set of wireless links between node pairs. Further, consider three consecutive nodes of the WMN \emph{viz.}, $x_s$, $x_d$ and $l_s$, which lie on the positive \textit{X-axis} at a distance of $x_s$, $x_d$ and $l_s$ from the origin, respectively. The protocol model deems a packet transmission on link x ($x_s$ to $x_d$) successful, if and only if  $\forall$  $l$ $\in$ $E - \{x\}$, we have
 \begin {equation}
  \lvert l_s \textendash x_d \rvert \hspace{1mm} \geq (1 + \varDelta) \hspace{1mm} \lvert x_s \textendash x_d \rvert \hspace{5mm}  and
 \end {equation}
 \begin{equation}
  \lvert  x_s \textendash x_d \rvert \hspace{1mm} \leq \hspace{1mm} R_c  
 \end{equation}
Where: 
\begin{itemize}
 \item \textit{$(x_s,0)$ is the source of link x.}
 \item \textit{$(x_d,0)$ is the destination of link x.}
 \item \textit{$(l_s,0)$ is the source of other link $l$ whose destination is $(x_d,0)$.}
 \item \textit{$\varDelta$ is a positive parameter.}
 \item \textit{$R_c$  stands for the effective range of communication.}
\end{itemize}
\item \textbf{\textit{The Interference Range Model}} :  The model mandates the spatial separation between a receiver and an arbitrary interferer to be greater than a fixed quantity \emph{i.e.,} the interference range, for successful transmission. It can be considered to be a simplification of the Protocol model. 
\end{enumerate}
\subsection{Selection Of Interference Model}
The Protocol Model is a simplified representation of physical interference which we use in our work for three reasons. It is a simple yet felicitous mathematical representation of the actual wireless interference. It permits a binary interference modeling, \emph{i.e.,} a successful transmission is one which is not attenuated by any interfering signal active in its transmission range. 
Finally, there is no fixed interference range by which two communicating nodes need to be separated. Instead, the model gives us the flexibility of fixing the interference range which is proportional to the distance between a communicating node pair.

\subsection{Representing Interference In A WMN}
Having successfully identified the interfering wireless links in a WMN, we need to graphically represent their interference relationships. This representation is done by a special graph, called the \textit{Conflict Graph}. We now state a few concepts and definitions. Let $G=(V,E)$ represent an arbitrary WMN.
\begin{enumerate}
 \item \textbf{\textit{Potential Interference Link}} :
Let $i\in V$, $j\in V$, such that $(i,j) \in E$, then $\forall (m,n) \in E$, where the transmitting range of the radio at node $m$ or $n$, extends upto, or beyond node $i$ or $j$, are called the potential interference links of link $(i,j)$. They are also termed as conflicting links or contention edges. 
 \item \textbf{\textit{Potential Interference Number}}: 
Let $i \in V$, $j \in V$, then the potential interference number of link $(i,j) \in E$, is the total number of links in E which are the potential interference links of $(i,j)$. It is also called as the \textit{Interference Degree}.
\item \textbf{\textit{Total Interference Degree (TID)}} :  It is an approximate estimate of the adverse impact of the interference endemic in a WMN. It is arrived at by halving the sum of the \textit{potential interference numbers} of all the links in the graph.
 \item \textbf{\textit{Conflict Graph ($CG$)}} : $G_c=(V_c,E_c)$ is generated from graph $G=(V,E)$ where  
\begin{itemize}
 \item 
   $V_c = E$  or   $V_c$ = \{ $(i,j) \in E$  $\lvert$ $(i,j)$ \text { is a wireless communication link} \}
 \item \{ $((i,j), (m,n)) \in  E_c$  $\lvert$ $(m,n)  $ is a potential interference or conflict link of $(i,j)$ in $G$ \}.
\end{itemize}
\end{enumerate}
Thus the edges in graph G become the vertices in $G_c$. There exists an edge between two vertices $x_c \in V_c$ and $y_c \in V_c$, where $x_c=(i,j) \in E$ and $y_c=(m,n) \in E$, iff  the corresponding links in edge set $E$ of graph $G$ \emph{i.e.,} $(i,j)$ and $(m,n)$ are conflicting links. In other words, the wireless communication links in the WMN become the vertices in the conflict graph, and any two of these vertices share an edge iff the corresponding wireless links in the WMN interfere with each other. 

\section{Related Research Work} 
Interference substantially degrades the wireless network performance. It leads to low end-to-end throughputs, high packet loss and high transmission delays. Multi-hop transmissions in WMNs are adversely impacted by the co-channel interference, which deteriorates network capacity and destabilizes fairness in link utilization \cite{3Saadawi}. Addressing the issue of internal co-channel interference in a WMN is of foremost importance in WMN deployments. While designing network topology, efforts are made to limit the impact of interference by optimal node placement. An MRMC architecture is better suited to be interference resilient as compared to simpler architectures. Significant research work has been carried out towards alleviating the inimical impact of interference on WMN performance. For example, several MAC protocols for WMNs have been proposed \cite{7Nova}, nodes equipped with high-power directional antennas have been deployed~\cite{10Wang}, etc. But these solutions limit the scalability and span 
of WMNs, and are not practically viable. Thus, the most crucial design choices to ensure interference minimization in  WMNs without limiting their scalability are : channel assignment (CA) to radios, link scheduling and routing. Numerous CA schemes \cite{14Weisheng}, and a multitude of routing algorithms \cite{ranirout} have been contributed in the effort to mitigate and restrain the impact of interference in WMNs.

Conflict graphs serve as the primary indispensable tool for addressing various design and performance issues in WMNs. They are extensively used for modeling and estimating the interference degree in wireless and cellular networks \cite{22Ramachandran}. However, a basic conflict graph (CG) is only suited to a wireless network in which each node is equipped with a single radio. In order to model the interference in an MRMC WMN, the concept of $CG$ needs to be extended to an enhanced version called the \textit{multi-radio multi-channel conflict graph} or an \textit{MMCG}. Several research endeavors \cite{13Jorge, 14Weisheng, 15Skalli, 16Subramanian, 17Xutao, 18Marina, 19Cao, 20Hongkun, 21Hamed, 22Ramachandran, 23Cheng, 24Aizaz}, directed at finding an efficient CA scheme for an MRMC WMN have made use of the concept of MMCG to model the interference in their network scenario.\\
 In \cite{15Skalli}, the authors merely suggest that the conflict graph was generated by ensuring that the \textit{interference-to-communication} ratio is set to $2$, with no further insight into the algorithmic aspects of this crucial step. In contrast, the literature in \cite{13Jorge} defines two approaches to generate conflict graphs. The first definition is centered on the \textit{traffic flow interference}, employing the protocol model and assuming unidirectional traffic flows. The second approach takes into account the \textit{link interference} based on the extended protocol model. However, its evident that neither of the proposed techniques is inherently generic in its outlook. The former assumes a unidirectional traffic flow and mandates the application of the protocol model as the underlying interference model, while the latter necessitates the use of the extended protocol model, restricting both their approaches to specific WMN architectures. Likewise, authors in \cite{
14Weisheng} define a conflict graph to be an undirected graph under the protocol model.\\
 Authors in \cite{16Subramanian} provide a high level definition of CG, with an assurance that the concept is applicable to any interference model. However, they do not explicitly propose any algorithm. Further, they opine that a CG does not change with the assignment of channels to vertices. This is not a true characteristic of the MMCG of a WMN, as it often changes when different CA schemes are deployed in the WMN. Assignment of different channels to a link, under different CAs, alters the set of its conflict links, thereby generating a different MMCG for each CA. Thus the authors' contention that a CG for a WMN will not change, does not hold true, at least in the context of an MRMC deployment.\\
 Authors in \cite{17Xutao} discuss a single-channel CG and its multi-channel peer (MMCG) for the protocol model, without suggesting a methodical approach to generate either. In addition, they map a link to a unidirectional flow, and a bidirectional traffic is represented by two links. This underscores the fact that researchers tend to perceive the interference dynamics which are a common feature of all WMNs, in a personalized way specific to their WMN model, rather than adopting a broadly applicable view. In \cite{18Marina}, the authors lay a special emphasis on \textit{weighted} conflict graphs for the protocol model. For non-interfering or non-overlapping channels, they recommend that a CG should be generated for each individual channel, and the CG for the WMN will be the union of all single channel CGs. Thus they take a single-channel view of an MRMC WMN, applied to multiple channels, and the final MMCG is an aggregate of the unique individual single channel CGs. The approach is intuitive and simple, but 
for a medium to large scale WMN where each node is equipped with multiple identical NICs, and which leverages the availability of a high number of non-interfering channels, this fragmented view of a WMN to arrive at an MMCG will cause substantial implementation overhead.

 The human perception of interference scenarios plays a significant role in the generation of a conflict graph. For example, the representation of WMN conflict relationships in the MMCG proposed in \cite{20Hongkun} is quite different than the one suggested in \cite{21Hamed}. The work in \cite{20Hongkun} aims to create a multi-dimensional CG by making use of a radio-link-channel tuple, while authors in \cite{21Hamed} create a link-layer flow contention graph which is essentially a simple CG, that is based upon the number of channels allocated and the channel assignment to interfaces. Based on our review, we opine that the most generic and widely applicable of all MMCG creation procedures is suggested by \textit{Ramachandran et. al.} \cite{22Ramachandran}. The authors extend the conflict graph concept to model a multi-radio WMN, and generate a \textit{multi-radio conflict graph} or an \textit{MCG}. They describe the MCG generation approach explaining the use an improvised vertex coloring algorithm to color the 
MCG, which ensures that each radio in the network is assigned a single channel. Yet, a lucid algorithm to generate the MCG is not proposed in the study. Any correlation between the MCG creation approach and the underlying interference model is not presented either. This leaves room for ambiguity in determining the dependence of the MCG creation technique upon the interference model being employed. Most importantly, the suggested approach and the corresponding illustration does not address the interference caused and experienced by SCRs operating on identical frequencies, and thus fails to represent the RCI scenarios in its interference estimate. From the literature review presented above, we can conclude that the underlying fundamental concepts of a conflict link and a basic conflict graph, have been rightly employed by researchers in their work. However, the creation of an MMCG in the research studies often depends upon one or more of the following factors.
 \begin{enumerate}
  \item The Interference Model being used, which is the \textit{protocol model} in most studies.
  \item The WMN Topology
  \item Representation of Traffic Flows \emph{i.e.,} unidirectional or bidirectional
  \item Perception of the interference scenarios
 \end{enumerate}
 
  Further, we have not come across any research study related to MMCG generation that adequately addresses or even highlights, the detrimental effect of multiple SCRs installed at a WMN node, which have been assigned the same channel to communicate on. In the upcoming sections we investigate and delve into the phenomena of RCI.
  
 \section{Spatial Co-location Of Radios \& Radio Co-location Interference}
An important aspect which most of the existing MMCG creation techniques fail to acknowledge, is the effect of spatial co-location of radios on wireless links, emanating from a node equipped with multiple radios. Such a node certainly stands to benefit if each one of its radios is assigned a different RF channel and can concurrently communicate with adjacent nodes, substantially raising the capacity of the node and the entire network. The throughput at a node and by virtue of aggregation, the capacity of the entire network,  can be tremendously accentuated if the RF channels being assigned are non-interfering or orthogonal. However, if two or more SCRs are operating on the same RF channel (or overlapping channels), the multi-radio deployment is rendered ineffective and its advantages are negated. It is also adversely impacted by the additional radio co-location interference generated due to the close proximity of these SCRs. We restrict our study of RCI to SCRs operating on the same channel, which is 
consistent with the binary interference model that we have adopted. In this section, we investigate the impact of spatial co-location of radios on the overall interference dynamics in a WMN. We elucidate two RCI scenarios to support the theoretical proposition, and then experimentally validate the suggested argument. 
\subsection{Two Radio Co-location Interference Scenarios}
\subsubsection{Case 1}
\begin{figure}[htb!]
                \centering
                \includegraphics[width=8.5cm, height=1.8cm]{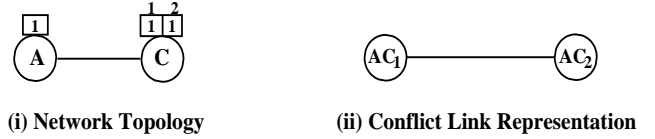}
                \caption{Impact of spatially co-located radios}
                \label{Co-Location1}
        \end{figure}

Consider a trivial two node wireless network illustrated in Figure~\ref{Co-Location1}. Node $A$ is equipped with one IEEE 802.11g radio while node $C$ is equipped with two identical IEEE 802.11g radios \emph{i.e.,} a pair of identical SCRs. Nodes $A$ and $C$ lie within each others transmission range and thus share a wireless communication link in the WMN. 

 Since the wireless communication link between the nodes $A$ and $C$ is a RF transmission, by virtue of RF wave propagation, any transmission from node A will reach both radio $C_1$ and radio $C_2$  alike, as they are co-located at $C$. Similarly, both radio $C_1$ and radio $C_2$ are independently capable of a simultaneous transmission to the radio on node $A$. In the above scenario, links $AC_1$ and $AC_2$ are interfering links and ought to have an edge in the corresponding MMCG to represent their mutual conflict.
 
    \begin{figure}[htb!]
                \centering
                \includegraphics[width=8.5cm, height=7cm]{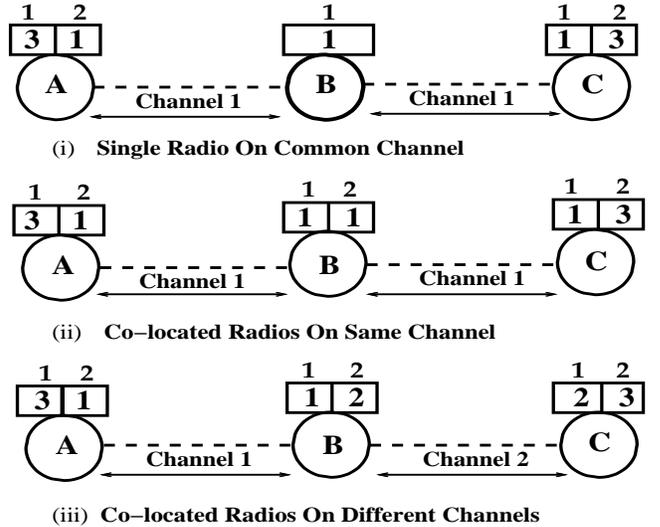}
                \caption{Scenarios of radio co-location }
                \label{Co-Location2}
    \end{figure}
    \subsubsection{Case 2}
    Consider the three flavors of a WMN layout illustrated in the three scenarios of Figure~\ref{Co-Location2}. The nodes are equipped with one or more IEEE 802.11g radios, which are operating on the depicted radio frequencies. One of the three orthogonal channels $1$, $2$, \& $3$, as per the 802.11g specifications, are allocated to the radios. 
    In Figure~\ref{Co-Location2}~$(i)$, the extreme nodes $A$ and $C$ are equipped with a pair of SCRs, while the node at the center $B$, has a single radio. In Figure~\ref{Co-Location2}~$(ii)$, all the three nodes are equipped with a pair of SCRs. Both the scenarios employ common-channel communication, which is evident from the radio-channel allocations. There is, however, a fundamental difference between the two scenarios with respect to node $B$. Figure~\ref{Co-Location2}~$(i)$ emulates a \textit{Single-Radio} architecture, rejecting any possibility of RCI. In contrast, Figure~\ref{Co-Location2}~$(ii)$ has dual radios at node $B$, both are assigned the same channel, and thus B becomes the epicenter of RCI. The wireless links $AB$ and $BC$ in the depicted WMN layout are the potential conflicting links. Let us focus on node  $B$ and reiterate the argument presented in \textbf{Case 1} above. For the layout in Figure~\ref{Co-Location2} $(i)$, only the radio-links $A_2B_1$ \& $C_1B_1$ are conflicting links. 
But for the layout in Figure~\ref{Co-Location2}~$(ii)$, the  total interference degree escalates substantially as there are six conflicting radio-link pairs, viz. $A_2B_1$ \& $A_2B_2$, $A_2B_1$ \& $B_1C_1$, $A_2B_1$ \& $B_2C_1$, $A_2B_2$ \& $B_1C_1$, $B_2C_1$ \& $B_1C_1$ and $A_2B_2$ \& $B_2C_1$. In the WMN layout in Figure~\ref{Co-Location2}~$(iii)$ the wireless links $AB$ and $BC$ are represented by the radio-links $A_2B_1$ and $B_2C_1$. They have been assigned channels $1$ and $2$, respectively, and therefore are non-conflicting links. This scenario fully utilizes the inherent multi-radio multi-channel architecture, leading to an interference free deployment.

  Drawing from the stated theoretical arguments we contend that in the considered trivial WMN layout, the \textit{Single Radio Common Channel} (SRCC) operation in Figure~\ref{Co-Location2}~$(i)$ will perform better, even if marginally, than the \textit{Multi Radio Common Channel} (MRCC) operation in Figure~\ref{Co-Location2}~$(ii)$. Further, the \textit{Multi Radio Different Channel} or \textit{MRDC} deployment in  Figure~\ref{Co-Location2}~$(iii)$ will significantly outperform the other two configurations.  
  
 \subsection{ Experimental Validation}
To corroborate our argument with actual experimental data, simulations of the three network layouts illustrated in Figure~\ref{Co-Location2} were performed in ns-3 \cite{NS-3}.
\subsubsection{Network Design} 
The inbuilt ns-3 TCP BulkSendApplication is employed to establish two TCP connections between node pairs $A$ \& $B$ and $B$ \& $C$, for the scenarios $(i)$, $(ii)$, and $(iii)$ of Figure~\ref{Co-Location2}. We install a TCP \textit{BulkSendApplication} source at the nodes $A$ \& $C$. Both the TCP sinks are installed at node $B$, as it is the node common to all the conflicting links, and hence the focal point of maximum interference offered to the data communication. Every source sends a \textit{10 MB} data file to its corresponding sink. We observe the \textit{Network Aggregate Throughput} for each of the three scenarios, which serves as a fair metric to gauge the adverse impact of interference in the WMN layout. Simulation parameters are listed in Table~\ref{CoLocationSimParam}.
\begin{table}[ht]
\caption{Co-location Experiment Simulation Parameters}
\begin{tabular}{|p{4cm}| p{4cm}|}
\hline
\bfseries
 Parameter&\bfseries Value \\ [0.2ex]
 \hline
\hline
 Transmitted File Size &10 MB  \\
\hline
 Maximum 802.11 Phy Datarate &9 Mbps  \\
\hline
RTS/CTS&Enabled  \\
\hline
TCP Packet Size &1024 Bytes    \\
\hline
Fragmentation Threshold&2200 Bytes  \\
\hline
Inter-node Separation &250 mts\\
\hline
Propagation Loss Model&Range Propagation  \\
\hline
\end{tabular}
\label{CoLocationSimParam}
\end{table}   

\subsubsection{Simulation Results}
For each deployment, multiple independent sample runs were performed. 
We register the throughput of only the onward flow, \emph{i.e.,} the \textit{Source - Sink} flow, for both TCP connections of each deployment, and illustrate the \textit{Average Aggregate Network Throughput} recorded for the three scenarios, in Table~\ref{CoLocationResults}.

\begin{table}[ht]
\caption{Co-location Experiment Simulation Results}
\center
\begin{tabular}{|p{4.1cm}| p{1cm}| p{1cm}|p{1cm}|}
\hline
\bfseries     Parameter&\bfseries SRCC&\bfseries MRCC&\bfseries MRDC \\ [0.5ex]
\hline
Average Aggregate Network Throughput (Mbps)&\bfseries 3.19 &\bfseries 3.01&\bfseries 5.87  \\
\hline
\end{tabular}
\label{CoLocationResults}
\end{table}  

The \textit{SRCC} deployment performs slightly better than the \textit{MRCC} deployment. This result vindicates our theoretical contention, that the surge in interference due to SCRs operating on a common channel, degrades the network performance substantially. Here, the difference in the \textit{Average Network Aggregate Throughput} of the two deployments is less than $10\%$, which is not remarkable.
However, we should consider the fact that in medium to large MRMC WMNs, the presence of SCRs which have been assigned the same channel is in moderate to large numbers. This increase in SCRs will exacerbate the adverse effects of RCI on the network performance. The \textit{Multi-Radio Different-Channel} or the \textit{MRDC} deployment offers an \textit{Aggregate Network Throughput}, that is almost twice that of a trivial \textit{SRCC} deployment. This is in conformity with our theoretical supposition as well. We can safely conclude, that the results of this investigation substantiate the proposed theoretical concept of RCI \emph{i.e.,} the \textit{interference caused by SCRs} which have been assigned a common-channel. 
\subsection{Impact Of RCI On Network Performance}
The severity of RCI in a wireless network depends upon the number of SCRs that have been assigned an identical channel. The two factors which decide the impact of RCI are, the number of available orthogonal channels offered by the wireless technology and the number of radios that a node in a wireless network is equipped with. We now elaborate on these factors.
 \begin{itemize}
  \item \textbf{Number of orthogonal channels :~}Availability of orthogonal channels depends upon the wireless communication technology being used. For example, IEEE 802.11g radios operating in the 2.4 GHz spectrum have only 3 orthogonal channels at their disposal. Therefore, RCI has a significant impact on network performance as we have demonstrated in our study. However, 12 to 24 non-overlapping channels are available to the IEEE 802.11n/ac radios operating in the 5 GHz spectrum. This will ensure that SCRs are seldom assigned identical channels, and the impact of RCI is minimal.
  \item \textbf{Number of radios per node :~}The number of radios installed on a wireless node depends upon the size, cost and most importantly, the power or energy consumption of a radio. These three factors limit the number of radios that can be installed on a node.
 \end{itemize}
 There are two primary reasons why addressing the impact of RCI on WMNs is of great significance. First, IEEE 802.11g/n radios operating in the 2.4 GHz band are being used in \mbox{Wi-Fi} networks across the globe, especially in the developing countries. In this study, we will demonstrate the severity of impact that RCI has on medium to large WMNs employing the 802.11g standard through extensive simulations. Thus, RCI alleviation measures and techniques will immensely benefit these WMN/WLAN deployments. Second, in the IEEE 802.11n/ac networks operating in 5 GHz spectrum, currently the number of available orthogonal channels far exceeds the number of radios installed on a node. But in the foreseeable future, this relationship may change and even be reversed. Our argument is based on the developments made in the field of radio hardware in the last decade. Significant inroads have been made in the development of carbon nanotube based nanoradios \cite{nanorad}, first proposed by the well known physicist Alex 
Zettl. A communication architecture for nanoradios has been proposed in \cite{nanoarch} which provides an operational framework for the carbon nanotube based nanoradios.  Further, in \cite{nanoart} authors discuss the significance of current pioneering research in the field of nanoradios, and describe the role nanonetworks are poised to play in revolutionizing the modern communication networks. With the advent of nanoradio technology, it will not be unrealistic to imagine a single wireless device equipped with dozens of nanoradios. In addition, the challenge of energy constraints in radio communication will be met through innovations in the field of micro-power and nano-power systems \cite{micropower}. Taking these developments into consideration, it is possible that in future the number of radios present on a node in a WMN will be quite comparable to the number of available orthogonal channels. Thus, the impact of RCI in such wireless networks will be highly detrimental.

It is thus imperative that we take cognizance of the RCI phenomena, and address its adverse impact on the performance of a WMN. The first step of this exercise would be to account for and appropriately represent the co-location interference scenarios in the interference model of a WMN. But the representation of RCI is lacking in the MMCG creation method suggested in \cite{22Ramachandran}, and all other research works aimed at minimization of interference in WMNs. The underlying reason is that while creating the MMCG, the fact that multiple radios installed on the same node are spatially co-located is overlooked. A result of this oversight is that a few interference scenarios escape notice during the MMCG creation and the estimate is seldom a true reflection of the actual interference present in the network.  Our work adopts a fundamental approach towards RCI mitigation, which is through creation of radio co-location aware MMCGs. Since no prior research work focuses on the impact of RCI on WMNs and its 
mitigation, it makes our work relevant and indispensable.
Having laid the theoretical foundations, supplemented with experimental evidence, we formally state the research problem we aim to pursue and address in the next sub-section.

\subsection{Problem Definition}
\textit{Multi-Radio Multi-Channel conflict Graphs} or \textit{MMCGs} are frequently used to accurately represent the interference present in a WMN and measure its intensity or degree of impact on the WMN. Thus, a generic approach to create an MMCG for any arbitrary WMN is of utmost importance. The need for a comprehensive procedure to generate an MMCG $G_c =(V_c ,E_c)$ for a given input WMN graph $G=(V,E)$, which is independent of the factors such as the \textit{WMN topology}, the \textit{choice of interference model}, the \textit{channel allocation} scheme etc., is  often felt by researchers attempting to solve CA, routing or maximum-throughput problems in a WMN.\newline 
To the best of our knowledge, a lucid, all-encompassing and explicitly proposed algorithm for MMCG creation, especially one which factors in the effects of SCRs, is lacking in the current research literature. The novel concept of RCI and the alleviation of its adverse impact on the performance of a WMN, through adequate and accurate representation in the creation of its MMCG, is what distinguishes our study from the plethora of approaches suggested before. 

\section{The Proposed MMCG Generation Algorithms}

To remedy the lack of a generic algorithm, we now propose two comprehensive polynomial time algorithms to create a multi-radio multi-channel conflict graph or an \textit{MMCG}. The algorithms are designed with the vision of creating a widely applicable and versatile method of generating MMCGs. We ensure their structural and functional independence from the commonly encountered constraints listed below. 
\begin{enumerate}
 \item {\textbf{\textit{The Choice of Interference Model}}}: As stated earlier, we employ the protocol interference model to determine the interfering links. This however, is not binding upon the algorithms, and any other interference model can be chosen instead. The algorithms allow the underlying interference model to define a conflicting link. Thus the choice of the interference model is not an implementation constraint. 
 \item {\textbf{\textit{WMN Topology}}}: The algorithms are topology independent and applicable to all WMN deployments. The graphical representation, however, must be a connected graph. This condition is reasonable, necessary and not an impediment, as having an isolated node with no wireless connectivity to any other node in the WMN is wasteful.
 \item {\textbf{\textit{Number of Radios and Channels}}} : The algorithms are also applicable to both single channel and multi-channel WMN deployments \emph{i.e.,} they can create MMCGs for the same WMN topology for both, a common CA and a varying/multiple CA. 
 \item {\textbf{\textit{The Channel Assignment or CA Scheme }}}: The algorithms can generate an initial \textit{multi-radio conflict graph}, whose edges denote potentially interfering links when every radio is assigned the same channel. They can as easily generate a \textit{multi-radio multi-channel conflict graph} or an MMCG, which depicts the actual state of interference in a WMN deployment in where a CA scheme is implemented. The output MMCGs will certainly differ if the CA scheme implemented in the network changes.
 \item {\textbf{\textit{Interpretations of Interference Scenarios}}}: The algorithms, especially the one which takes RCI into account, consider every possible interference scenario in the WMN. Thus, they avoid the variations associated with the human interpretations of interference scenarios.
 \end{enumerate}

The first algorithm we propose, does not take into account the effect of spatial co-location of multiple radios on a node. It follows a conventional approach, variations of which have been employed in various research endeavors, customized to suit the model being implemented. However, a versatile algorithm which is independent of the aforementioned constraints has not been formally proposed, a void we now bridge.
We christen it \textit{The Classical MMCG} algorithm, or \textit{C-MMCG}. The second algorithm effectively factors the RCI into its interference modeling logic. It paints a more comprehensive and wholesome picture of the interference scenario in the given WMN, and is thus a notable improvement over \textit{C-MMCG}. We name it \textit{The Enhanced MMCG} algorithm, or \textit{E-MMCG} for ease of reference.
The two broad scenarios in a WMN where the proposed MMCG algorithms find great utility are elucidated below. 
\begin{enumerate}
 \item \textit{Prior to CA } : Before the CA exercise is carried out in a WMN, usually all the radios are assigned a default channel. The MMCG resulting from this default channel configuration represents a maximal prevalent interference scenario and thus serves as an ideal input to a CA algorithm. 
 \item \textit{After CA } :  After the radios have been assigned appropriate channels in accordance with the applied CA scheme, the proposed MMCG algorithms may be used to generate the TID estimate for the WMN. This desirable feature facilitates a theoretical assessment of the efficacy of the CA approach employed.
\end{enumerate}
Now we propose the two algorithms along with their functional description.

\renewcommand{\algorithmicrequire}{\textbf{Input:}}
\renewcommand{\algorithmicensure}{\textbf{Output:}}
\begin{algorithm}[htb!] 
\caption{C-MMCG  : Radio Co-location Not Considered}
\label{C-MMCG}
\begin{algorithmic}[1]
\REQUIRE $G = (V,E)$, $R_i (i \in V)$, $N_i(i \in V)$ = \{ $j \lvert (j \in V)$ \&\& $(i\neq j)$ \&\& $((i,j) \in E)$ \} \\
\textit{Initially} $:$ $V'\leftarrow \O{}$, $E'\leftarrow \O{}$, $V_c\leftarrow \O{}$, $E_c \leftarrow \O{}$\\
\textit{Notations} $:$ $G$ $\leftarrow$ WMN Graph, $R_i$ $\leftarrow$ Radio-Set, \mbox{$N_i$ $\leftarrow$ Neighbour Set}
\ENSURE $G_c$ $=$ $(V_c,E_c)$\\
\noindent\rule{8.2cm}{0.4pt}
\FOR {$ i \in V$}
\STATE $ V' \leftarrow V' + R_i$
\FOR {$ j \in N_i$}
\FOR {$ x \in R_i$}
\FOR {$ y \in R_j$}
\STATE $E' \leftarrow E' + (x,y)$
\ENDFOR
\ENDFOR
\ENDFOR
\ENDFOR
\COMMENT {Get the intermediate graph $G'=(V',E')$}
\FOR {$ (i,j) \in E'$  where  $i \in  V', j \in  V'$}
\STATE $V_c \leftarrow V_c + (i,j)$
\ENDFOR 
\COMMENT {Create the Vertex Set $V_c$ of the $CG$  $G_c$}
\FOR {$v \in V_c , u \in V_c , v \neq u$}
\STATE Use an Interference Model to determine if $u$  \&  $v$ are Potentially Interfering Links
\IF { True }
\IF {$ (Channel(u)== Channel(v))$}
\STATE $E_c \leftarrow E_c  + (u,v) $
\COMMENT {Create the Edge Set $E_c$ of the $CG$  $G_c$}
\ENDIF 
\ENDIF 
\ENDFOR	
\COMMENT {Output C-MMCG $G_c=(V_c,E_c)$}
\end{algorithmic}
\end{algorithm}

\subsection{The Classical MMCG Algorithm }
The \textit{C-MMCG} algorithm  adopts a conventional approach to model the interference endemic in WMNs.
Its stepwise procedure is described in Algorithm~\ref{C-MMCG}. Steps 1 to 10 split each node in the original WMN topology graph $G=(V,E)$, into the number of radios it is equipped with, and generate an intermediate graph $G'=(V',E')$, where $V'$ represents the set of total number of wireless radios in the WMN and $E$ is edge set of links between radio pairs.\\
While $G$ reflects a node centric view of the WMN, $G'$ reflects the view of the WMN at the granularity of individual radios. Step $2$ splits the radio set of each node in $G$ to individual radio-nodes in $G'$. Steps $3$ to $10$, process the neighbor set of a node in $G$ to create edges in $E'$, for each individual radio in the radio-set of the node in context. The intermediate graph $G'$ becomes the input for the final MMCG creation step. In steps $11$ to $13$, the vertex set $V_c$ of the MMCG is populated by adding elements of the edge-set $E'$.  Further, in steps $14$ to $21$, the vertices in $V_c$ are processed pairwise, and a corresponding edge is added to the MMCG edge-set $E_c$, \textit{iff} the vertex pair being currently processed is conflicting, and both the vertices have been assigned the same channel. The function \textit{Channel()} fetches the channel assigned to a particular vertex of $G_c$. As described earlier, the channel returned by the function would be the default channel if the algorithm 
is being applied to a WMN prior to the CA exercise. Else, \textit{Channel()} would fetch the channel that has been assigned by the CA scheme employed in the WMN. The underlying interference model determines whether a vertex pair is conflicting. We have employed the \textit{Protocol Interference Model}, but as stated earlier, any other interference model may be used as well. Algorithm~\ref{C-MMCG} finally outputs the \textit{C-MMCG}, $G_c$. The algorithmic time complexity is $O(n\textsuperscript2)$, as each of the three functional steps \emph{viz.,} creating the intermediate graph $G'$, generating the vertices of C-MMCG $G_c$, and finally adding the edge set to C-MMCG $G_c$, have an $O(n\textsuperscript2)$ computational complexity, where $n$ is the number of nodes in the WMN.

\subsection{The Enhanced MMCG Algorithm}
The \textit{E-MMCG} considers all possible interference scenarios that exist in a WMN, including the RCI. The stepwise procedure to generate an E-MMCG for a WMN is described in Algorithm~\ref{E-MMCG}. In addition to the \textit{C-MMCG} logic, Algorithm~\ref{E-MMCG} also captures interference due to spatial co-location of radios in the WMN. In steps $22$ to $28$, the algorithm adds an edge between two vertices of the E-MMCG, \textbf{iff}
\begin{enumerate}
 \item The corresponding pair of wireless links in the WMN originate or terminate at the same node  and,
 \item The links have been assigned the same channel. 
\end{enumerate}

\renewcommand{\algorithmicrequire}{\textbf{Input:}}
\renewcommand{\algorithmicensure}{\textbf{Output:}}
\begin{algorithm}
\caption{E-MMCG : Radio Co-location Considered}
\label{E-MMCG}
\begin{algorithmic}[1]
\REQUIRE $G = (V,E)$, $R_i (i \in V)$, $N_i(i \in V)$ = \{ $j \lvert (j \in V)$ \&\& $(i\neq j)$ \&\& $((i,j) \in E)$ \} \\
\textit{Initially} $:$ $V'\leftarrow \O{}$, $E'\leftarrow \O{}$, $V_c\leftarrow \O{}$, $E_c \leftarrow \O{}$\\
\textit{Notations} $:$ $G$ $\leftarrow$ WMN Graph, $R_i$ $\leftarrow$ Radio-Set, \mbox{$N_i$ $\leftarrow$ Neighbour Set} 
\ENSURE $G_c$ $=$ $(V_c,E_c)$\\
\noindent\rule{8.2cm}{0.4pt}
\FOR {$ i \in V$}
\STATE $ V' \leftarrow V' + R_i$
\FOR {$ j \in N_i$}
\FOR {$ x \in R_i$}
\FOR {$ y \in R_j$}
\STATE $E' \leftarrow E' + (x,y)$
\ENDFOR
\ENDFOR
\ENDFOR
\ENDFOR
\COMMENT {Get the intermediate graph $G'=(V',E')$}
\FOR {$ (i,j) \in E'$  where  $i \in  V', j \in  V'$}
\STATE $V_c \leftarrow V_c + (i,j)$
\ENDFOR 
\COMMENT {Create the Vertex Set $V_c$ of the $CG$  $G_c$}
\FOR {$v \in V_c , u \in V_c , v \neq u$}
\STATE Use an Interference Model to determine if $u  \&  v$ are Potentially Interfering Links
\IF { True }
\IF {$ (Channel(u)== Channel(v))$}
\STATE $E_c \leftarrow E_c  + (u,v) $
\COMMENT {Create the Edge Set $E_c$ of the $CG$  $G_c$}
\ENDIF 
\ENDIF 
\ENDFOR	
\FOR {$v \in V_c$, $u \in V_c$, $v \neq u$ ; $v = (a,b)$, $u = (c,d)$ ; \\$a,$ $b,$ $c,$ $d \in V'$}
\FOR  {$i \in V$}
\IF { $ [ \{ (a \in R_i\| b \in R_i)$ \&\& $(c \in R_i \| d \in R_i)\}$ \&\& (Both elements of $R_i$ on same channel)]}
\STATE $E_c \leftarrow E_c  + (u,v)$
\ENDIF
\ENDFOR
\ENDFOR
\COMMENT {Output E-MMCG $G_c=(V_c,E_c)$}
\end{algorithmic}
\end{algorithm}

The RCI accounting steps will apply to both common and multiple channel deployments in the WMN, preserving its generic nature. E-MMCG thus ensures that the interference scenarios discussed in Section $4$, which are not being addressed in the existing research literature are accounted for. Further, the injection of the RCI into the overall interference dynamics is duly represented by addition of necessary and sufficient links in the E-MMCG. The links added to the E-MMCG to account for the RCI, are characteristic of the E-MMCG algorithm, and are generated from its steps $22$ to $28$. These conflicting links may or may not be determined by the employed interference model, but they will definitely not escape notice of the E-MMCG algorithm. The time complexity of the algorithm, similar to its conventional counterpart C-MMCG is $O(n\textsuperscript2)$.

\subsection{C-MMCG and E-MMCG : An Illustration}
\begin{figure}[htb!]
	  \centering
	  \includegraphics[width=8.5cm, height=7cm]{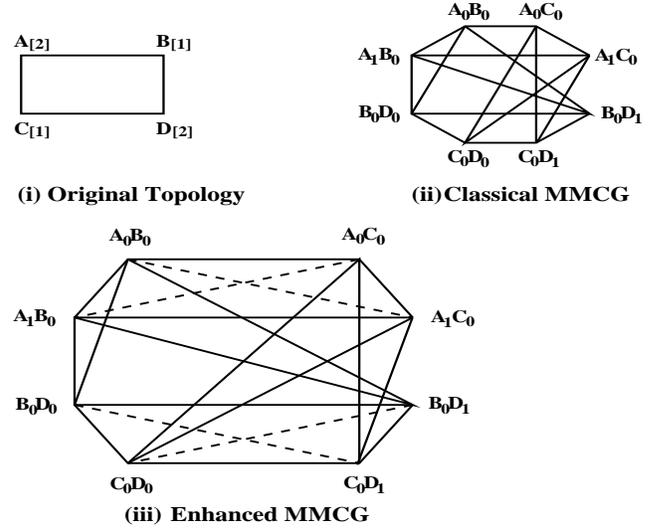}
	  \caption{An illustration of Classical and Enhanced MMCGs for a given Topology }
	  \label{MMCGFig}
  \end{figure}

Let us pictorially demonstrate the output MMCGs for the two flavors proposed above, through Figure~\ref{MMCGFig}. Figure~\ref{MMCGFig} $(i)$ depicts the original WMN topology comprising of four nodes, $A$, $B$, $C$ and $D$, where the nodes are assigned $2$, $1$, $1$, and $2$ number of radios, respectively. Each radio is operating on the default channel, so the two methods will generate the initial, maximal-conflict MMCG for the WMN. The graphical representations of C-MMCG and E-MMCG, for the given WMN layout, are exhibited in cases $(ii)$ and $(iii)$ of Figure~\ref{MMCGFig}, respectively. Upon observation, it is evident that E-MMCG has all the conflicting links present in C-MMCG, and in addition contains four more interfering links, \emph{viz.} $A_0B_0 - A_1C_0$, $A_1B_0 - A_0C_0$, $B_0D_0 - C_0D_1$ and $C_0D_0 - B_0D_1$. These four conflicting links are the result of RCI, which is caused by the wireless transmissions from radios spatially co-located at nodes $A$ and $D$.

 The number of these additional conflict links caused by RCI increases drastically with the size of the WMN, which we demonstrate in Section $6$. The ability of E-MMCG algorithm to capture and represent the interference scenarios spawned by RCI is the first step towards alleviating the adverse impact of RCI. 

\section{Simulations, Results and Analysis}  
Having proposed the two MMCG algorithms, it is imperative we prove their relevance to real-world WMN deployments. We take a three pronged approach in this regard.

\subsection{Measuring Impact Of Interference} \label{ssec1}

\begin{figure}[htb!]
               \centering
                \includegraphics[width=8.5cm, height=6cm]{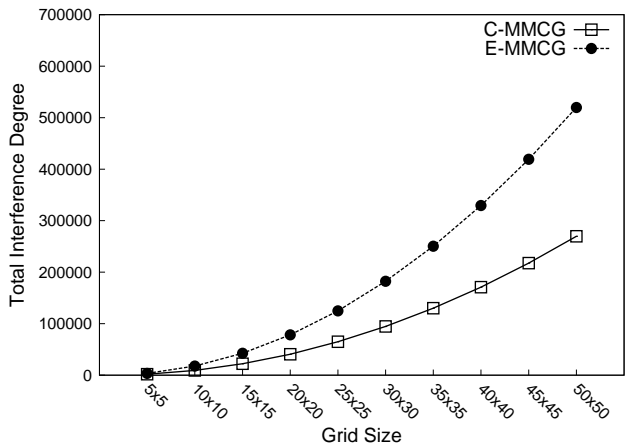}
                \caption{TID Comparison of C-MMCG vs E-MMCG}
                \label{C-MMCGvsE-MMCG}
        \end{figure}

 We employ the MMCG algorithms to measure the TID in a WMN, and compare the results of the two flavors.
 We consider a square \textit{Grid Layout} for the WMNs, of size $5n\times5n$ where $n=\{1,2,\ldots,10\}$. Thus the size of WMNs varies from $5\times5$ nodes to $50\times50$ nodes, where all the nodes are equipped with $2$ identical radios, and all radios are on a \textit{common channel}. This configuration represents a \textit{maximal interference} scenario, and is ideal for analysis. We apply both the MMCG algorithms to each of these grid topologies. The results illustrated in Figure~\ref{C-MMCGvsE-MMCG} underscore the fact that the E-MMCG algorithm accounts for all the \textit{potential interfering links} or \textit{interference scenarios} present in a WMN, and hence registers substantially high values of \textit{TID}.
 
 In contrast, the C-MMCG algorithm suffers from a limited potency to probe a WMN for \textit{potential interfering links} as it does not factor in the presence of SCRs operating on a common channel. This is reflected by its poor accounting of \textit{Interference Degree} values as compared to its enhanced counterpart, the E-MMCG.
Further, as the size of WMN grows, the difference in the \textit{TID} of the two MMCG approaches becomes increasingly prominent. This implies that there is a tremendous upsurge in the RCI as the size and complexity of the WMN increases. This finding further consolidates the proposition we put forward in Section $4$, that the adverse impact of RCI gets more pronounced in medium to large WMNs. 
        
\subsection{Application to CA Algorithms}        
 
Since \textit{Conflict Graphs} serve as the input to \textit{Channel Assignment} algorithms, the next logical step is to apply the MMCG algorithms to two graph-theoretic solutions of the CA problem. In \cite{22Ramachandran} authors propose a \textit{Breadth First Search} approach or BFS-CA, which is a centralized dynamic algorithm that employs the services of a \textit{channel assignment server} or CAS. Initially, the radios in the WMN are assigned a default channel which experiences the least interference from intentional or un-intentional interferers in close proximity, based on a channel-ranking technique. The CAS computes the average distance of each vertex in the multi-radio conflict graph (MCG) from the gateway. Thereafter, the algorithm performs a breadth-first scan of the MCG, starting from the vertices closest to the gateway, and assigns a channel to every vertex that it encounters. To every vertex, CAS tries to assign a channel which is orthogonal to the channels assigned to its neighboring nodes. 
Else, it selects a channel randomly from the set of available channels and allots it to the current vertex. 
A \textit{Maximal Independent Set} channel assignment scheme or MaIS-CA is proposed in \cite{24Aizaz}. It is a greedy heuristic scheme, which determines the maximal independent set of vertices in a conflict graph, assigns them an identical channel and then removes them from the conflict graph. This process is iterated, until all the vertices have been assigned a channel.\\
We opine that MaIS-CA is algorithmically superior to BFS-CA as its CA approach distributes the channels among the radios in a more balanced fashion, and also assures a higher degree of connectivity in the WMN graph. For a theoretical validation of the stated notion, we implement these two CA algorithms in Grid WMNs. Both C-MMCG and E-MMCG versions serve as the input to CA schemes, and the \textit{TID} for each CA scheme is determined. The nodes are equipped with 2 identical radios each, and we utilize the 3 non-overlapping channels guaranteed by IEEE 802.11g specifications. For a smooth discourse hereon, we adopt the following nomenclature to differentiate between the CAs.
\begin{itemize}
 \item \textit{C-MMCG based CAs :} BFS-CA$_1$ and MaIS-CA$_1$.
 \item \textit{E-MMCG based CAs :} BFS-CA$_2$ and MaIS-CA$_2$.
\end{itemize}

The procedure we follow is described below :
 \begin{itemize}
 \item Take WMN grid of size  \textit{n$\times$n}, where \textit{n $\in$ \{3,5,10\}}.
  \item Create two \textit{MMCGs} using the algorithms C-MMCG and E-MMCG.
  \item Use both flavors of MMCG as input to BFS-CA and MaIS-CA to obtain final CAs, $4$ in all. 
  \item Apply C-MMCG on BFS-CA$_1$ \& MaIS-CA$_1$ and E-MMCG on BFS-CA$_2$ \& MaIS-CA$_2$, to estimate their respective \textit{TIDs}.
  \end{itemize}
\par\noindent
This procedure will furnish the theoretical measure of the impact of interference in each of the final CAs. For consistency, we subject a particular version of a CA only to the corresponding MMCG algorithm, to compute the \textit{TID} estimate. Further, we only compare two CAs generated from the same MMCG approach. Comparing the interference estimate of a C-MMCG CA with an E-MMCG CA is not logical, because the approaches to generate these estimates are not identical.
 \begin{table} [h!]
\caption{A Comparison of TIDs of the MMCG CAs}
\center
\tabcolsep=0.11cm
\begin{tabular}{|p{1.6cm}|p{1.5cm}|p{1.5cm}|p{1.5cm}|p{1.5cm}|}
\hline 
     \multicolumn{1}{|c|}{} & \multicolumn{4}{|c|}{TID}\\  \cline{2-5}
   \multicolumn{1}{|c|}{Grid Size} &\multicolumn{2}{|c|}{BFS-CA} & \multicolumn{2}{|c|}{MaIS-CA}\\ \cline{2-5}
     & C-MMCG & E-MMCG  & C-MMCG & E-MMCG  \\
\hline  
$3 \times 3$&82&70&16&56\\
\hline  
$5 \times 5$&436&716&142&488\\
\hline  
$10 \times 10$&2098&2470&834&2036\\
\hline  
\end{tabular} 
\label{MMCGCA}
\end{table}

\noindent
It can be inferred from Table~\ref{MMCGCA}, that the BFS-CA of a particular MMCG version registers a higher measure of interference than the corresponding MaIS-CA, \emph{i.e.,} with respect to \textit{TID} of CAs, \textit{BFS-CA$_1$} $>$ \textit{MaIS-CA$_1$} and \textit{BFS-CA$_2$} $>$ \textit{MaIS-CA$_2$}. This result strengthens the argument that MaIS-CA is a better CA scheme than BFS-CA. Further, it offers a theoretical assurance that employing the use of E-MMCG approach does not alter the intrinsic algorithmic disposition of a CA.

\subsection{Simulation Testbed For Performance Evaluation Of CAs} {\label{sec}}
The final step in this research investigation entails that we monitor and analyze the performance of BFS-CA and MaIS-CA, for both versions of the MMCG, through comprehensive simulations. We create an extensive data traffic scenario by considering various single-hop and multi-hop transmission combinations. Our objectives are four fold.
 \begin{enumerate}
  \item To compare the performance characteristics of E-MMCG CA against that of the C-MMCG version, for the same CA algorithm.
  \item To compare the performance of the two approaches, BFS-CA and MaIS-CA, for both versions of MMCG.
  \item To observe the relative difference between the performances of BFS-CA and MaIS-CA, in the two versions of MMCG.
  \item To observe the traffic interruptions or abrupt flow terminations, for a CA, in both versions of MMCG.
 \end{enumerate}
 \noindent
Through objectives \textit{(b)} and \textit{(c)} above, we intend to study the consistency in CA performance \emph{i.e.,} if \textit{X-CA} performs better than \textit{Y-CA} in the C-MMCG model, then we opine that it should outperform \textit{Y-CA} in the E-MMCG model as well. Further, we study the difference between the performances of \textit{X-CA} and \textit{Y-CA} for a relative comparison.
\subsubsection{Simulation Design Parameters}
We consider a $5\times5$ grid WMN, which provides some semblance of a large-scale topology. To gauge the impact of interference on the WMN, we choose the \textit{Aggregate Throughput} of the network, as the primary performance metric. The total capacity of a network consistently degrades with the increase in interference, and it is thus a suitable and sufficient metric. We also employ \textit{Packet Loss Ratio} and \textit{Mean Delay} of the network as metrics for some scenarios. While Packet Loss Ratio gives a measure of data packets lost during communication, the Mean Delay provides the end-to-end latency in packet transmission. Thus both of them are reliable indicators of the disruption caused in data transmission by the prevalent interference.

TCP and UDP are the underlying transport layer protocols being used in the experiments. The inbuilt ns-3 models of \textit{BulkSendApplication} and \textit{UdpClientServer} are utilized for TCP and UDP implementations, respectively. TCP simulations are aimed at estimating the \textit{Aggregate Network Throughput} while the UDP simulations are employed to determine the \textit{Packet Loss Ratio} and the \textit{Mean Delay}. For ease of reference, hereon we denote the three metrics as Throughput$_{Net}$, PLR and MD, respectively.
\par\noindent
The radios installed on all nodes are identical IEEE 802.11g radios, operating in the standard specified \mbox{$2.4$ $GHz$} spectrum that offers up to 14 channels, of which 3 channels are orthogonal. We restrict the number of available channels to these 3 non-interfering channels. We employ the \textit{ERP-OFDM} modulation technique, with a ceiling of $9$ $Mbps$ on the maximum $PHY$ data-rate. We let the transmission power assume the default value of $16.02$ $dBm$, and set the receiver gain to $-10$ $dBm$ for better sensitivity. Nodes are placed at a separation of $200$ $mts$, so that the adjacent nodes lie comfortably within each other's transmission range. Use of \textit{Range Propagation Loss Model} in ns-3 facilitates an easy implementation of the \textit{protocol model} for interference modeling. The remaining simulation parameters are listed in the Table~\ref{FinalSim}.
 \begin{table} [h!]
\caption{ns-3 Simulation Parameters}
   \center 
\begin{tabular}{|p{5cm}|p{3cm}|}
\hline
\bfseries
 Parameter&\bfseries Value \\ [0.2ex]
 \hline
\hline
Grid Size&$5\times5$   \\
\hline
No. of Radios/Node&2   \\
\hline
Range Of Radios&250 mts   \\
\hline
Available Orthogonal Channels&3   \\
\hline
Maximum 802.11 PHY Datarate &9 Mbps  \\
\hline
Maximum Segment Size (TCP)&1 KB   \\
\hline
Packet Size (UDP)&1KB\\
\hline
Fragmentation Threshold&2200 Bytes  \\
\hline
RTS/CTS(TCP) &Enabled  \\
\hline
RTS/CTS(UDP) &Disabled  \\
\hline
Routing Protocol Used &OLSR    \\
\hline
Loss Model&Range Propagation   \\
\hline
Propagation Model&Constant Speed   \\
\hline
\end{tabular}
\label{FinalSim}
\end{table}   

\subsubsection{Data Traffic Characteristics}
\begin{figure}[htb!]
   \raggedleft
      \includegraphics[width=8cm, height=6cm]{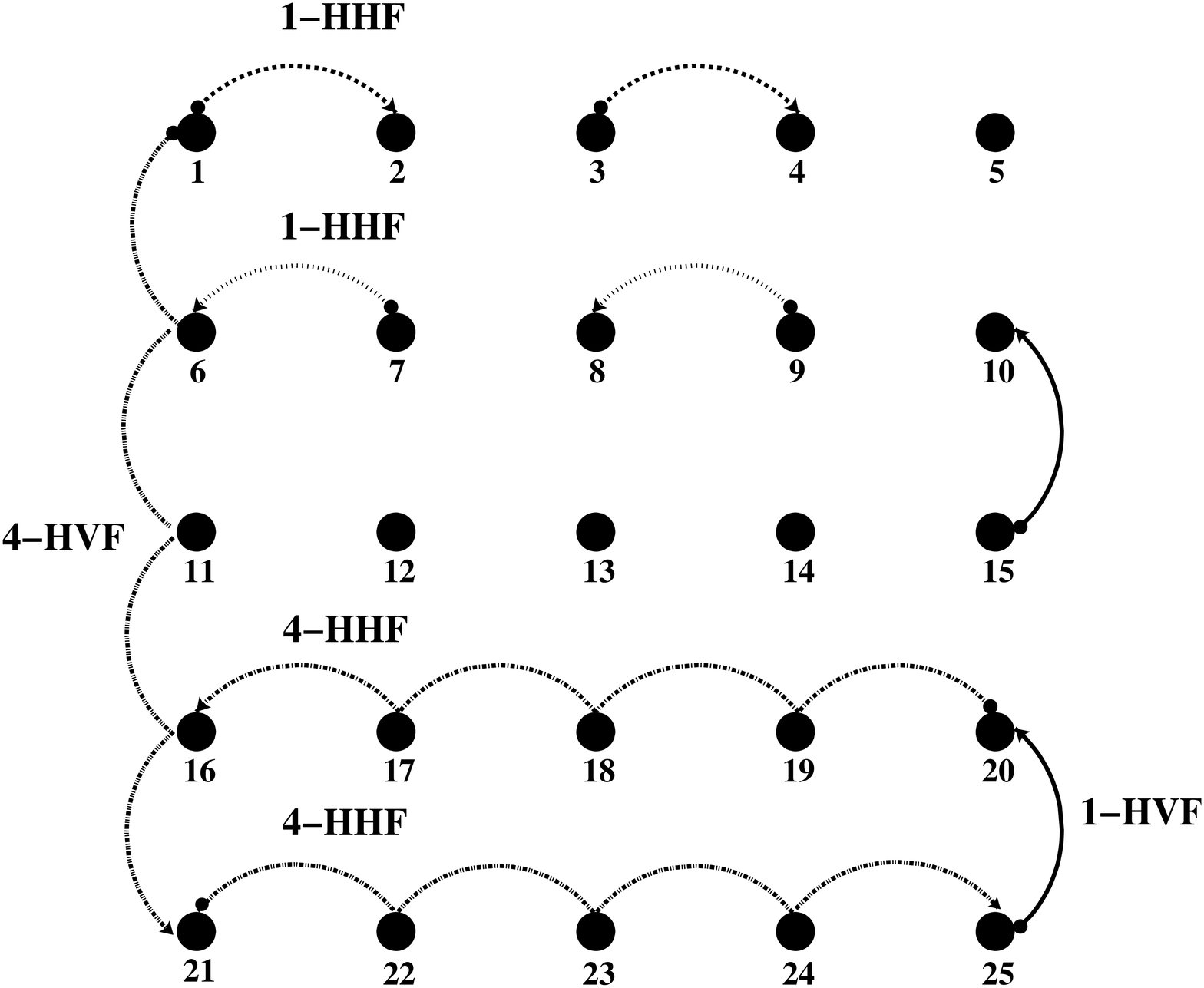}
      \caption{Grid Layout}
      \label{Grid}
\end{figure}

The most critical step in studying the impact of interference in a WMN is to tailor the right set of traffic flows, which will identify and expose the interference bottlenecks. To simulate a data traffic with suitable characteristics, we consider five types of TCP/UDP traffic flows which include both, single and multi-hop flows. We deploy a combination of these flow-types to characterize the intensity of interference present in a $5\times5$ grid WMN. The 25 nodes in the WMN grid are numbered from 1 to 25, for the sake of representation. The traffic flows are depicted in Figure~\ref{Grid}, followed by a brief description of each flow. The TCP/UDP client or \textit{source}, can be identified by the dotted tail of the link representing the TCP/UDP connection, while the arrow-head signifies the TCP/UDP server or the \textit{sink}. 

 \noindent
 \begin{enumerate}
  \item \textit{One Hop Horizontal Flow or 1-HHF} : Single Hop TCP connections are established between alternate node pairs in all the rows of the WMN grid depicted in Figure~\ref{Grid}. For example, in row 1, node-pairs (1 \& 2) and (3 \& 4) have a one hop TCP connection. TCP source application is installed on the node represented by a smaller number, and the sink application on the node bearing the bigger number in the node-pair.
  \item \textit{One Hop Vertical Flow or 1-HVF} : In addition to the 1-hop horizontal flows, one hop TCP vertical flows, between alternate node-pairs in each column in the bottom-up direction, are also generated. 
  \item \textit{Four Hop Horizontal Flow or 4-HHF} : Since multi-hop transmission is an inherent trait of WMNs, we deploy TCP/UDP connections between the first and last nodes of each row, which are four hops away.
  \item \textit{Four Hop Vertical Flow or 4-HVF} : To capture the spatial interference characteristics of the grid in the vertical direction, TCP/UDP connections are established in a top-down fashion \emph{i.e.,} between the first and last nodes of each column, which are four hops away. 
  \item \textit{Eight Hop Diagonal Flows or 8-HDF} : The diagonally opposite node-pairs ($1$ \& $25$) and ($21$ \& $5$) have a TCP/UDP connection each, and generate eight-hop TCP/UDP flows. This is the maximum possible hop-length between a source and destination pair, in the given simulation grid. 
  
  \end{enumerate}
\subsubsection{Simulation Terminology, Scenarios and Statistics} 
To keep the discourse lucid and coherent, we first define some terms we use in the upcoming sections. 
 \begin{enumerate}
  \item \textit{Flow} : It refers to the onward TCP/UDP traffic flow from a TCP/UDP source to the TCP/UDP sink.
  \item \textit{Abrupt Flow} : In all the TCP connections, we mandate that the source should transmit \textit{10 MB} data to the sink. For a TCP connection to be deemed successful, it is imperative that the sink should receive the \textit{10 MB} data sent by the source in entirety. Else, the TCP connection is considered to have abruptly terminated, and the flow to be an \textit{Abrupt Flow}. Abrupt flows are a measure of the obstructions caused by prevalent interference to the data transmissions in a WMN. This notion is predicated on the fact that routing failures are often caused by high levels of interference in multi-hop wireless networks \cite{Abrupt}. Loss of routing information causes packets buffered at intermediate relay nodes to be dropped. Since the routing protocol is singularly responsible for the routing mechanism, a TCP source may never be aware of an alternate route or route re-establishment. Thus, after subsequent failed attempts at re-transmitting the packets, a connection-timeout is invoked 
at the source and eventually the flow is abruptly terminated.
  \item \textit{Abrupt Flow Count} (AFC) : It is the total number of Abrupt Flows encountered in all the simulations of a \textit{Test Case Class}.
  \item \textit{Throughput$_{Net}$} : It refers to the \textit{Network Aggregate Throughput}, which is the aggregate throughput of all the flows in a simulation.
  \item \textit{Flow-RX} : A 4-HHF TCP flow in any row \textit{X} of the grid.
  \item \textit{Flow Type-Y or FT-Y} : A set of all possible combinations of 4-HHFs taken \textit{Y} at a time, where \textit{Y $\in$ \{1...5\}}. Thus, FT-1 would be a set containing  \textsuperscript5$C_1$ or five 4-HHFs, viz. Flow-R1, Flow-R2, Flow-R3, Flow-R4 and Flow-R5.
   \end{enumerate}

We segregate the simulation scenarios into combinations of one-hop flows and multi-hop flows. The underlying motivation is to monitor the behavior of CAs for single-hop flows, and more complex multi-hop flows, separately. The test-cases have been categorized into the following three classes. 
   \begin{enumerate}
   \item \textbf{Test Case Class 1 : Flow Sustenance Testing }\\
   In the test-cases belonging to this class, numerous one-hop TCP connections are concurrently active. The motivation here is to highlight the capability of a WMN to establish and sustain multiple TCP connections under the debilitating effects of interference. Thus, these test-cases are focused on the number of abrupt flows encountered \emph{i.e.}, the AFC, rather than the Throughput$_{Net}$. The test-cases under this class are listed below.
   \begin{enumerate}
   \item \textit{Test Case 1} : Only vertical flows \emph{i.e.,} 1-HVFs.
   \item \textit{Test Case 2} : Only horizontal flows \emph{i.e.,} 1-HHFs.
   \item \textit{Test Case 3} : All vertical and horizontal flows \emph{i.e.,} 1-HVFs + 1-HHFs.
   \end{enumerate}
   \item \textbf{Test Case Class 2 : Flow Injection Testing}\\
   Multi-hop transmissions are a primary characteristic of WMNs. Concurrent multi-hop data connections are the prefect instruments for estimating the adverse impact of interference on the network capacity, as they transmit in tandem, triggering and intensifying the intricate interference bottlenecks in a WMN. We start with a single 4-HHF, and then inject one additional 4-HHF in each subsequent test-case. We monitor the network response to injection of fresh four-hop flows in terms of the network capacity, for a variety of 4HHF combinations. Here we focus only on monitoring the throughput response of the network and not the number of abrupt flow terminations. \\
   The goal here is to record the variation in network performance with the continuous injection of additional four-hop flows, hence it will suffice to do so for flows along the rows of the grid. We consider the five 4-HHFs \emph{i.e.,} Flow-R1...Flow-R5, and create a test-case for each \textit{Flow Type-Y} or \textit{FT-Y}, where \textit{Y $\in$ \{1...5\}}. Thus, we have the following test-cases.
   \begin{enumerate}
   \item \textit{Test Case 1} :   FT-1 \emph{i.e.,} \textsuperscript5$C_1$ 4-HHF combinations.
   \item \textit{Test Case 2} :   FT-2 \emph{i.e.,} \textsuperscript5$C_2$ 4-HHF combinations.
   \item \textit{Test Case 3} :   FT-3 \emph{i.e.,} \textsuperscript5$C_3$ 4-HHF combinations.
   \item \textit{Test Case 4} :   FT-4 \emph{i.e.,} \textsuperscript5$C_4$ 4-HHF combinations.
   \item \textit{Test Case 5} :   FT-5 \emph{i.e.,} \textsuperscript5$C_5$ 4-HHF combinations.
   \end{enumerate}
   \item \textbf{Test Case Class 3 : Load or Stress Testing}\\
   A reliable measure of network performance is often gauged under peak load, as it exhibits a network's resilience to bottlenecks that occur only at high traffic demands. We perform four test-cases of increasing data traffic demands. By virtue of the rise in the number of concurrent radio transmissions, there is an increase in the interference complexities of the network. For each of these scenarios, we observe and analyze not only the network capacity, but also the packet loss ratio (PLR) and mean delay (MD). Thus both TCP and UDP simulations are run for each of the the test-cases described below. 
   \begin{enumerate}
   \item \textit{Test Case 1} :   D2   \emph{i.e.,} Concurrent twin diagonal TCP/UDP flows or 8-HDFs.
   \item \textit{Test Case 2} :   H4V4 \emph{i.e.,} Eight Concurrent TCP/UDP flows comprising of adjacent 4-HHF and 4-HVFs, each taken four at a time. A total of four such combinations, for which simulations are run and the average Throughput$_{Net}$ is considered.
   \item \textit{Test Case 3} :   H5V5 \emph{i.e.,} Ten concurrent TCP/UDP flows consisting of all five 4-HHFs and all five 4-HVFs.
   \item \textit{Test Case 4} :   H5V5D2 \emph{i.e.,} Twelve concurrent TCP/UDP flows consisting of all five 4-HHFs, all five 4-HVFs, and both 8-HDFs. 
   \end{enumerate}
   \end{enumerate} 
\subsection{Results and Analysis}
The four CAs \emph{viz.}, BFS-CA$_1$, BFS-CA$_2$, MaIS-CA$_1$ and MaIS-CA$_2$, are subjected to the test-cases described above. The metrics we monitor and register for subsequent analysis are \textit{Throughput$_{Net}$}, \textit{AFC}, \textit{PLR} and \textit{MD}. We now present the recorded results, and methodically analyze them in adherence to the four objectives stated in sub-section $6.3$.

 \begin{figure}[htb!]
\center 
      \includegraphics[width=8.5cm, height=6cm]{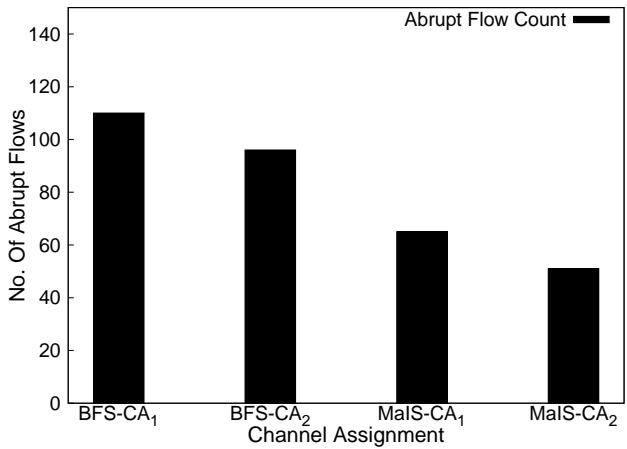}
      \caption{Abrupt Flow Count in Test Case Class 1 }
      \label{TC1}
\end{figure}
\subsubsection{Test Case Class 1}\mbox{} \\ 
The metric of relevance here is the \textit{Abrupt Flow Count} or AFC, which for C-MMCG CAs is quite higher than the corresponding E-MMCG CAs. This can be inferred from the graph displayed in Figure~\ref{TC1}. BFS-CA$_2$ is able to achieve a $12.7\%$ reduction in abrupt termination of flows over BFS-CA$_1$. Likewise, we record a $21.53\%$ drop in AFC for MaIS-CA$_2$, when compared to MaIS-CA$_1$. Further, a comparison of the two CA approaches underscores the consistency of MaIS-CA in outperforming BFS-CA, for both the MMCG approaches. MaIS-CA$_1$ registers $40.9\%$ lesser abrupt flow terminations than BFS-CA$_1$, and this improvement is more accentuated in  MaIS-CA$_2$ where the frequency of abrupt flows depreciates by $46.8\%$ when compared to BFS-CA$_2$.\\
From the perspective of relative performance in terms of Abrupt Flow Count, E-MMCG approach heightens the edge that MaIS-CA has over BFS-CA. This is evident from the \textit{relative decrease} in AFC of $14.4\%$ achieved in the E-MMCG CAs, over their C-MMCG peers. This result is arrived at by the simple expression \textit{(\% Drop in E-MMCG $-$ \% Drop in C-MMCG) $/$ \% Drop in C-MMCG)}.Thus, the E-MMCG CAs not only reduce the abrupt terminations of flows when compared to the respective C-MMCG versions, but also enhance the performance of a better CA scheme (MaIS-CA) when compared to a less efficient algorithm (BFS-CA). 

Throughput$_{Net}$ is not an ideal metric to compare the CA performances in \textit{Test Case Class 1}, but for consistency the Throughput$_{Net}$ results are presented in Table~\ref{TC}. MaIS-CA performs unarguably better than BFS-CA which is in accordance with the TID estimates shown in Table~\ref{MMCGCA}. In all the test-cases E-MMCG CAs perform slightly better than their C-MMCG peers. The only exception is observed in \textit{Test Case 2} \emph{i.e.,} Only 1-HHFs, where MaIS-CA$_1$ $>$ MaIS-CA$_2$. This reversal is rectified in \textit{Test Case 3} \emph{i.e.,} All 1-HHFs and 1-HVFs, which is a more comprehensive test scenario and a relatively better case for observing Throughput$_{Net}$.
\begin{table} [h!]
\center
\caption{Throughput$_{Net}$ in Test Case Class 1}
\tabcolsep=0.11cm
\renewcommand{\arraystretch}{1.1}
\begin{tabular}{|p{2cm}|p{2cm}|p{2cm}|p{2cm}|} 
\hline
  \multicolumn{4}{|c|}{\textbf{Network Throughput (Mbps)}}\\
  \hline
   \multicolumn{2}{|c|}{\textbf{BFS-CA}} &\multicolumn{2}{|c|}{\textbf{MaIS}} \\
   \hline
    \hline
    \multicolumn{4}{|c|}{\textbf{Test Case 1 $\rightarrow$ 1-HVFs}}\\
    \hline 
    \textit{BFS-CA$_1$} & \textit{BFS-CA$_2$}  & \textit{MaIS-CA$_1$} & \textit{MaIS-CA$_2$}
     \renewcommand{\arraystretch}{1}\\
\hline  
16.73&18.40&24.31&24.44
\renewcommand{\arraystretch}{1.1}\\
\hline 
\hline
 \multicolumn{4}{|c|}{\textbf{Test Case 2 $\rightarrow$ 1-HHFs}}\\
    \hline 
    \textit{BFS-CA$_1$} & \textit{BFS-CA$_2$}  & \textit{MaIS-CA$_1$} & \textit{MaIS-CA$_2$}
     \renewcommand{\arraystretch}{1}\\
\hline
19.05&20.41&21.95&19.55
\renewcommand{\arraystretch}{1.1}\\
\hline
\hline
\multicolumn{4}{|c|}{\textbf{Test Case 3 $\rightarrow$ 1-HVFs + 1-HHFs}}\\
    \hline 
      \textit{BFS-CA$_1$} & \textit{BFS-CA$_2$}  & \textit{MaIS-CA$_1$} & \textit{MaIS-CA$_2$}
     \renewcommand{\arraystretch}{1}\\
\hline 
24.99&25.36&32.35&34.16\\
\hline 
\end{tabular}
\label{TC}
\end{table}

 \begin{figure}[htb!]
\center 
\includegraphics[width=8.5cm, height=6cm]{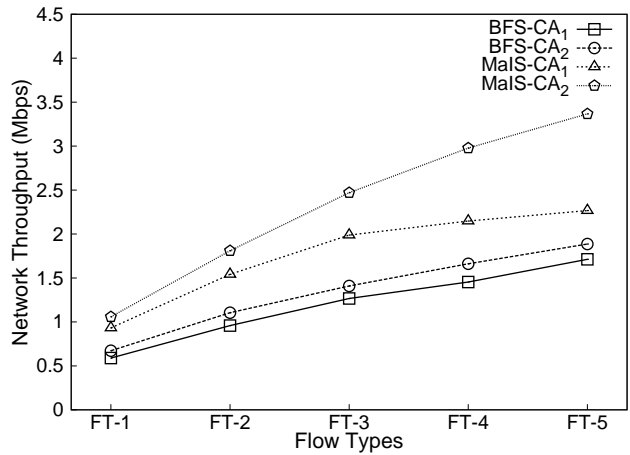}
\caption{Throughput$_{Net}$ in Test Case Class 2}
\label{TC2}
\end{figure}
\subsubsection{Test Case Class 2}\mbox{} \\ 
We now analyze the response of the CA deployments in the WMN in terms of observed Throughput$_{Net}$, as new four-hop flows are injected in the network. Results of  \textit{Test Case Class 2}, which consists of a variety of multi-hop test-cases, are exhibited as a graph in Figure~\ref{TC2}. The Throughput$_{Net}$ recorded for each of the five test-cases in this class, represented by \textit{FT-Y}, where \textit{Y $\in$ \{1...5\}}, is plotted for the four CA schemes. 
It is clearly evident that the E-MMCG version of a CA outperforms the C-MMCG version by a significant margin. For a reference, we quote the statistics of FT-5 from Figure~\ref{TC2}, for all the four CAs. The Throughput$_{Net}$ values of BFS-CA$_1$, BFS-CA$_2$, MaIS-CA$_1$, and  MaIS-CA$_2$, are recorded to be \textit{1.711 Mbps}, \textit{1.88 Mbps}, \textit{2.26 Mbps} and \textit{3.36 Mbps}, respectively.  We process the results to determine the change, \emph{i.e.,} increase or decrease, in the observed Throughput$_{Net}$ values of the two variants of the same CA, in Table~\ref{TC2T1}. The Throughput$_{Net}$ value of the C-MMCG version of a CA is considered as the base. 
\begin{table} [h!]
\caption{ \% Change in Throughput$_{Net}$ values of an E-MMCG CA over corresponding C-MMCG CA in Test Case Class 2}
\raggedleft
\tabcolsep=0.11cm
\begin{tabular}{|M{1.8cm}|M{1cm}|M{1.1cm}|M{1.1cm}|M{1.1cm}|M{1.1cm}|}
\hline 
    \multicolumn{1}{|c|}{\textbf{CA}} & \multicolumn{5}{|c|}{\textbf{\% Change in Throughput$_{Net}$ in FT}}\\  \cline{2-6}
    \multicolumn{1}{|c|}{\textbf{Strategy}}&\textbf{1}&\textbf{2}&\textbf{3}&\textbf{4}&\textbf{5}\\
\hline  
BFS&13.9&15.2&11.2&14.3&10.2\\
\hline  
MaIS&13.4&17.4&24.2&38.7&48.5\\
\hline  
\end{tabular} 
\label{TC2T1}
\end{table}

Throughput$_{Net}$ values of BFS-CA$_2$ are higher than those of BFS-CA$_1$ for all Flow Types, however always within the modest range of  $10\%$ to $15.2\%$. A more prominent increase in Throughput$_{Net}$ can be noticed in MaIS-CA$_2$ with respect to MaIS-CA$_1$. The rise in Throughput$_{Net}$ values ranges from $13.4\%$ in FT-1, to a maximum of $48.5\%$ in FT-5. This increase in Throughput$_{Net}$ of MaIS-CA$_2$, continues to rise from Flow Type-1 to Flow Type-5, \emph{i.e.,} with the increase in the number of concurrent flows injected in the network. 
The second objective is to assess how the two CA schemes fare against one another, in both the MMCG models. It can be inferred that MaIS-CA performs substantially better than BFS-CA, irrespective of the MMCG model. However, it is of great relevance to study the variation of the difference in Throughput$_{Net}$ values recorded for the two CA schemes, in the two MMCG models. Thus, we compute the \% difference in Throughput$_{Net}$ values of BFS-CA and MaIS-CA for each MMCG model in Table~\ref{TC2T2}. Throughput$_{Net}$ values of BFS-CA are considered as the base, and the \% increase or decrease of MaIS-CA over BFS-CA is calculated, for the particular MMCG variant. A \% decrease, is preceded by a negative sign.
\begin{table} [h!]
\caption{ \% Difference in Throughput$_{Net}$ values of BFS-CA and MaIS-CA for an MMCG approach in Test Case Class 2}
\raggedleft
\tabcolsep=0.11cm
\begin{tabular}{|M{1.8cm}|M{1cm}|M{1.1cm}|M{1.1cm}|M{1.1cm}|M{1cm}|}
\hline 
   \multicolumn{1}{|c|}{\textbf{MMCG}} & \multicolumn{5}{|c|}{\textbf{\% Difference in Throughput$_{Net}$ in FT}}\\  \cline{2-6}
   \multicolumn{1}{|c|}{\textbf{Model}}&\textbf{1}&\textbf{2}&\textbf{3}&\textbf{4}&\textbf{5}\\
\hline  
C$-$MMCG&58.1&60.7&56.8&47.7&32.3\\
\hline  
E$-$MMCG&57.4&63.7&75.3&79.2&78.4\\
\hline  
\hline 
Relative Difference ($\%$) &\textbf{-1.2}&\textbf{4.9}&\textbf{32.5}&\textbf{66}&\textbf{142}\\
\hline 
\end{tabular} 
\label{TC2T2}
\end{table}

In the C-MMCG deployment, MaIS-CA$_1$ records a significant increase over BFS-CA$_1$ that falls in the range of $32.3\%$ to $60.7\%$. In the E-MMCG deployment, MaIS-CA$_2$ surpasses its C-MMCG variant, registering tremendous increase over BFS-CA$_2$ Throughput$_{Net}$ is in the range of $57.4\%$ to $79.2\%$. Further, we calculate the \textit{relative difference} of the increase in Throughput$_{Net}$ that MaIS-CA shows over BFS-CA, between the two MMCG models, as a \% with C-MMCG as the base. The relative difference is slightly negative at ($-1.2\%$) for FT-1, but this result is not unsettling for two simple reasons. First, that FT-1 is a minimalistic test scenario with just one 4-HHF active at a time, and second being the diminished magnitude of this relative decrease. Besides, the \% relative increase rises immensely from FT-2 through FT-5, to reach a high of $142\%$ at FT-5, which is the most comprehensive interference scenario.\\

 \begin{figure}[htb!]
\center 
\includegraphics[width=8.5cm, height=6cm]{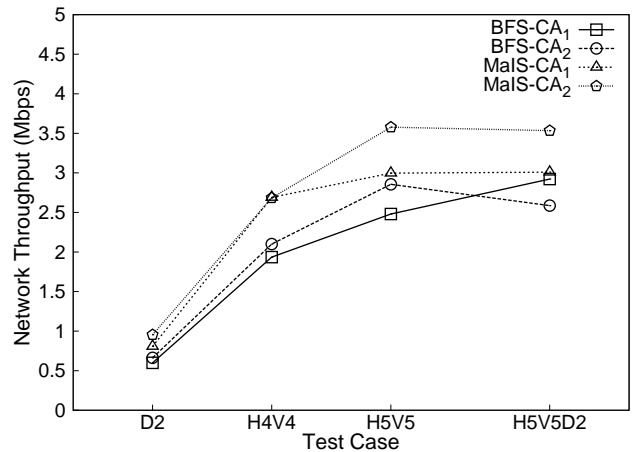}
\caption{Throughput$_{Net}$ in Test Case Class 3}
\label{TC3}
\end{figure}

\subsubsection{Test Case Class 3}\mbox{} \\ 
This class of test-cases is aimed at measuring the network performance in terms of Throughput$_{Net}$, PLR and MD, under heavy network data traffic. The test-cases of this class are complex as they involve a high number of concurrent multi-hop TCP/UDP connections, causing almost every interference scenario to affect the data transmission. The Throughput$_{Net}$ results of the \textit{stress testing} exercise are presented for analysis in the graphs depicted in Figure~\ref{TC3}. 
The overall performance of E-MMCG CAs continues to be better than their corresponding C-MMCG peers, but we can observe a few deviations from the trend. In test-case H4V4, MaIS-CA$_1$ registers a higher Throughput$_{Net}$ than MaIS-CA$_2$ although by an insignificant margin of $0.24\%$. Further, in the peak load scenario H5V5D2, the Throughput$_{Net}$ of BFS-CA$_1$ is higher than that of BFS-CA$_2$ by $11.4\%$, which can be noticed in Table~\ref{TC3T1}. The two instances in which a C-MMCG CA performs equal to or better than the corresponding E-MMCG CA do not raise any doubts about the efficacy of the E-MMCG model, but instead highlight the temporal and spatial characteristics of the endemic interference. Even a reasonably good CA scheme may register a sub-par performance for a particular traffic scenario. In all the remaining test-cases, the E-MMCG CAs outperform their C-MMCG counterparts, registering noticeable increase in Throughput$_{Net}$ that falls within the range of $8.4\%$ to $19.3\%$.

\begin{table} [h!]
\caption{ \% Change in Throughput$_{Net}$ values of an E-MMCG CA over corresponding C-MMCG CA in Test Case Class 3}
\center
\tabcolsep=0.11cm
\begin{tabular}{|M{1.5cm}|M{1.2cm}|M{1.5cm}|M{1.5cm}|M{1.5cm}|}
\hline 
    \multicolumn{1}{|c|}{\textbf{CA}} & \multicolumn{4}{|c|}{\textbf{\% Change in Throughput$_{Net}$ in Test Case}}\\  \cline{2-5}
    \multicolumn{1}{|c|}{\textbf{Strategy}}&\textbf{D2}&\textbf{H4V4}&\textbf{H5V5}&\textbf{H5V5D2}\\
\hline  
BFS&10.4&8.4&15.1&-11.4\\
\hline  
MaIS&18.2&-0.2&19.3&17.4\\
\hline  
\end{tabular} 
\label{TC3T1}
\end{table}
Let us now examine how the two CA schemes fare against one another, in the two MMCG models. In Table~\ref{TC3T2}, the \% difference in Throughput$_{Net}$ values of BFS-CA and MaIS-CA for each MMCG model is computed. 
MaIS-CA proves to be better than BFS-CA, regardless of the MMCG model employed. Secondly, the \% difference between the E-MMCG CAs is more pronounced in all scenarios except for the test-case H4V4, where it is $27.8\%$ while it is $39\%$  for the C-MMCG model. This reversal is the outcome of the fact that both versions of MaIS-CA \emph{viz.} MaIS-CA$_1$ and MaIS-CA$_2$, demonstrate similar Throughput$_{Net}$ characteristics in H4V4, while BFS-CA$_2$ registers a higher value than BFS-CA$_1$ as expected. Thus, the \% relative difference of the increase in Throughput$_{Net}$ that MaIS-CA shows over BFS-CA, between the two MMCG models, is positive for all test scenarios except for test-case H4V4.
\begin{table} [h!]
\caption{ \% Difference in Throughput$_{Net}$ values of BFS-CA and MaIS-CA for an MMCG approach in Test Case Class 3}
\center
\tabcolsep=0.11cm
\begin{tabular}{|M{1.7cm}|M{1.2cm}|M{1.5cm}|M{1.5cm}|M{1.5cm}|}
\hline 
\multicolumn{1}{|c|}{\textbf{MMCG}} & \multicolumn{4}{|c|}{\textbf{\% Difference in Throughput$_{Net}$}}\\ 
    \multicolumn{1}{|c|}{\textbf{Model}} & \multicolumn{4}{|c|}{\textbf{in Test Case}}\\
    \cline{2-5}
    \multicolumn{1}{|c|}{}&\textbf{D2}&\textbf{H4V4}&\textbf{H5V5}&\textbf{H5V5D2}\\
\hline  
C$-$MMCG&34.6&39.0&20.8&3.0\\
\hline  
E$-$MMCG&44.0&27.8&25.3&36.6\\
\hline  
\hline 
Relative Difference ($\%$) &\textbf{27.2}&\textbf{-28.7}&\textbf{21.6}&\textbf{1124.1}\\
\hline 
\end{tabular} 
\label{TC3T2}
\end{table}

 \begin{figure}[htb!]
\center 
\includegraphics[width=8.5cm, height=6cm]{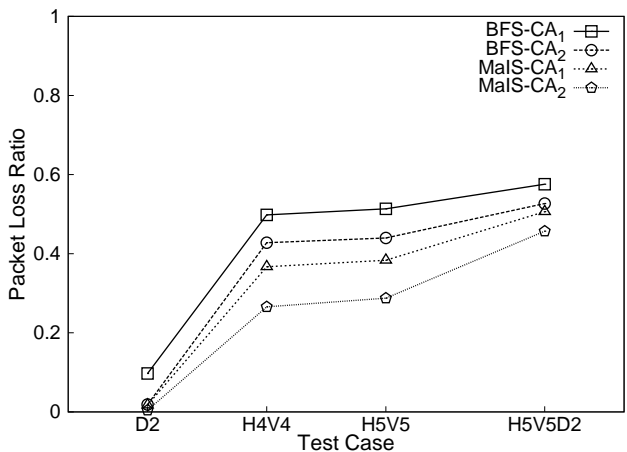}
\caption{Packet Loss Ratio in Test Case Class 3}
\label{TC3P}
\end{figure}

Moving on to the observed PLR values, let us examine the graph in Figure~\ref{TC3P} and the corresponding processed results in Table~\ref{TC3T3}. In all test-cases, the E-MMCG CAs suffer from a much lesser PLR then the respective C-MMCG CAs, thereby implying a reduced impact of interference on the data transmissions. Further, in accordance with the earlier result trends, PLR results also highlight that  MaIS-CA$_2$ registers a greater reduction in PLR over MaIS-CA$_1$ than BFS-CA$_2$ does over BFS-CA$_1$. 
\begin{table} [h!]
\caption{ \% Reduction in Packet Loss Ratio of an E-MMCG CA over corresponding C-MMCG CA in Test Case Class 3}
\center
\tabcolsep=0.11cm
\begin{tabular}{|M{1.5cm}|M{1.2cm}|M{1.5cm}|M{1.5cm}|M{1.5cm}|}
\hline 
    \multicolumn{1}{|c|}{\textbf{CA}} & \multicolumn{4}{|c|}{\textbf{\% Reduction in PLR in Test Case}}\\  \cline{2-5}
    \multicolumn{1}{|c|}{\textbf{Strategy}}&\textbf{D2}&\textbf{H4V4}&\textbf{H5V5}&\textbf{H5V5D2}\\
\hline  
BFS&80.7&14.1&14.3&8.4\\
\hline  
MaIS&75.5&27.5&25.0&9.7\\
\hline  
\end{tabular} 
\label{TC3T3}
\end{table}

\begin{table} [h!]
\caption{ \% Difference in Packet Loss Ratio between BFS-CA and MaIS-CA for an MMCG approach in Test Case Class 3}
\center
\tabcolsep=0.11cm
\begin{tabular}{|M{1.5cm}|M{1.2cm}|M{1.5cm}|M{1.5cm}|M{1.5cm}|}
\hline 
    \multicolumn{1}{|c|}{\textbf{MMCG}} & \multicolumn{4}{|c|}{\textbf{\% Difference in PLR in Test Case}}\\  \cline{2-5}
    \multicolumn{1}{|c|}{\textbf{Model}}&\textbf{D2}&\textbf{H4V4}&\textbf{H5V5}&\textbf{H5V5D2}\\
\hline  
C$-$MMCG&82.0&26.3&25.3&12.0\\
\hline  
E$-$MMCG&77.1&37.8&34.6&13.2\\
\hline  
\hline 
Relative Difference ($\%$) &\textbf{-5.9}&\textbf{43.7}&\textbf{36.7}&\textbf{10.1}\\
\hline 
\end{tabular} 
\label{TC3T4}
\end{table}
The \% relative difference of decrease in PLR values that MaIS-CA shows over BFS-CA, between the two MMCG models, is positive for all values except D2 where it is $-5.9\%$. For the remaining test-cases the relative difference is positive, always above $10\%$ and as high as $43\%$. Thus we can infer that the E-MMCG model accentuates the decrease in PLR registered by MaIS-CA over BFS-CA, as compared to the C-MMCG model where this decrease is less prominent.

\begin{figure}[htb!]
\center 
\includegraphics[width=8.5cm, height=6cm]{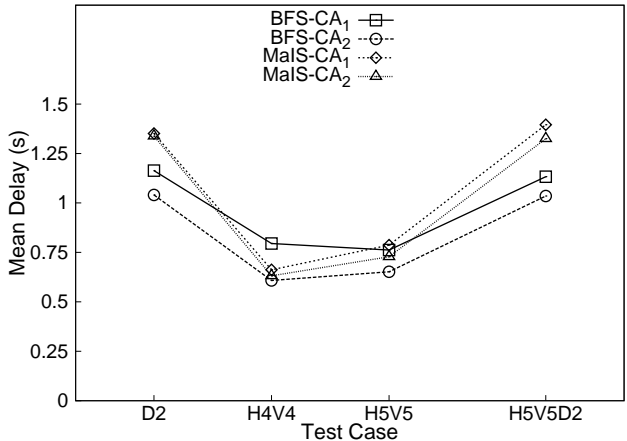}
\caption{Mean Delay in Test Case Class 3}
\label{TC3MD}
\end{figure}

The final metric of interest here is the mean delay (MD), the recorded results for which are depicted in the graph in Figure~\ref{TC3MD}. Deviating from the pattern of other metrics, MaIS-CA does not command a clear advantage over BFS-CA with respect to MD. In fact, the observed values for the two CAs fluctuate and can not be compared, which is easily discernible from Figure~\ref{TC3MD}. For most test-scenarios BFS-CA$_2$ registers the minimum MD values performing better than even MaIS-CA$_2$. Therefor, we restrict our analysis to the reduction in MD that an E-MMCG CA registers over its corresponding C-MMCG CA. The processed results are presented in Table~\ref{TC3T5}.
Both MaIS-CA$_2$ and BFS-CA$_2$ boast of a reduced MD than MaIS-CA$_1$ and BFS-CA$_1$, respectively. The difference between the two versions of BFS-CA is more pronounced which is a shift from the result patterns observed earlier. Nevertheless, the E-MMCG CAs succeed in reducing packet transmission latency in the WMN grid as compared to their conventional counterparts.
\begin{table} [h!]
\caption{ \% Reduction in Mean Delay of an E-MMCG CA over corresponding C-MMCG CA in Test Case Class 3}
\center
\tabcolsep=0.11cm
\begin{tabular}{|M{1.5cm}|M{1.2cm}|M{1.5cm}|M{1.5cm}|M{1.5cm}|}
\hline 
    \multicolumn{1}{|c|}{\textbf{CA}} & \multicolumn{4}{|c|}{\textbf{\% Reduction in MD in Test Case}}\\  \cline{2-5}
    \multicolumn{1}{|c|}{\textbf{Strategy}}&\textbf{D2}&\textbf{H4V4}&\textbf{H5V5}&\textbf{H5V5D2}\\
\hline  
BFS&10.5&23.5&14.4&8.6\\
\hline  
MaIS&1.0&4.5&7.4&5.0\\
\hline  
\end{tabular} 
\label{TC3T5}
\end{table}

Through these experiments we have successfully demonstrated that the same CA scheme exhibits better performance when it receives an E-MMCG as input, as compared to a C-MMCG as input. The E-MMCG ensures that a lesser number of SCRs are assigned identical channels by the CA scheme, which in turn leads to a decrease in RCI in the network. Further, the proposed MMCG algorithms enable a CA scheme to reduce the link conflicts and minimize the impact of interference, but do not make any assumptions or place any restrictions on the functioning of the CA schemes. Thus, the MMCG algorithms do not alter the algorithmic disposition of a CA scheme. 

Having stated the results of the testing effort, we now make reasonable inferences and logical conclusions. But before that, we make an observation regarding \textit{TID} in the next section.

\section{Total Interference Degree : A Reliable Theoretical Metric ?}
  The theoretical measure of \textit{Interference Degree} can be both \textit{local} \emph{i.e.,} of an individual node, and \textit{total}  \emph{i.e.,} of a CA scheme in entirety \cite{TID1}. This concept has been generously used in the WMN research literature to efficiently solve research problems such as the CA problem, the routing and scheduling problems etc \cite{TID2}. With respect to the CA problem, the guiding idea is that a CA with lesser TID, will be more efficient and register better performance as compared to a CA with a higher TID \cite{Ding}.  However, we have observed that this theoretical idea does not always hold true when compared to the actual experimental data. In any domain relying primarily on experimentation and actual deployments, the proposed theories are seldom accurate or precise. Thus there is an acceptable threshold of deviation of real-time performance from the theoretical predictions. But in the case of TID of CAs, we contend that the reasonable threshold of acceptance is 
breached. In this section, we elaborate upon this problem.
 
\begin{figure}[htb!]
\center 
\includegraphics[width=9cm, height=6cm]{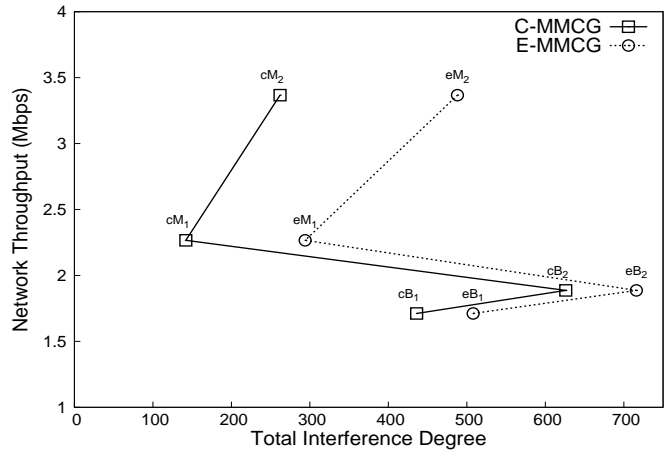}
\caption{Correlation of TID with Throughput$_{Net}$ for FT-5}
\label{ID2}
\end{figure}

 On the basis of recorded network metrics, we now arrange the four CAs in the following sequence of decreasing performance~:  MaIS-CA$_2$ $>$  MaIS-CA$_1$ $>$ BFS-CA$_2$ $>$  BFS-CA$_1$. Further, let us observe Table~\ref{MMCGCA} again, and consider the TID values of CAs for the $5 \times 5$ WMN grid. Earlier we had compared the TIDs of two CAs belonging to the same MMCG model and made theoretical inferences. Instead of that, now if we consider the TID values of all four CAs irrespective of the MMCG model, the theoretical order of expected performance based on the TID values would be~: MaIS-CA$_1$ $>$ BFS-CA$_1$ $>$  MaIS-CA$_2$ $>$  BFS-CA$_2$. It is clear that this theoretical sequence is not in conformity with the established experimental sequence. A counter-argument may be presented that since a particular MMCG model was employed to compute the TIDs only for the CAs of the same model, a comparison of cross-model TID values is not logical. Hence, for a consistency in the approach taken to generate TID 
values, we employ the following two alternative methods to generate the TID values for all four CAs, regardless of the MMCG model they belong to. 
 \begin{enumerate}
  \item Compute TIDs using the C-MMCG algorithm.
  \item Compute TIDs using the E-MMCG algorithm.
 \end{enumerate}
We plot the computed TID values against the Throughput$_{Net}$ values recorded in test-case FT-5 of \textit{Test Case Class 2}, in Figure~\ref{ID2}. The plot-lines for titles \textit{C-MMCG} and \textit{E-MMCG} correspond to the TIDs generated using the C-MMCG and the E-MMCG  algorithm, respectively. The \textit{(TID, Throughput$_{Net}$)} co-ordinates for the CA quartet MaIS-CA$_1$, MaIS-CA$_2$, BFS-CA$_1$, BFS-CA$_2$ are labeled as \textit{M$_1$, M$_2$, B$_1$} and \textit{B$_2$}, respectively. Further, the C-MMCG plot labels are prefixed by a \textit{'c'}, and the E-MMCG plot labels are prefixed by an \textit{'e'}.

\par\noindent
The plots for both the MMCG models are almost identical in shape and gradient, differing only in terms of the TID values. The E-MMCG CA versions register higher value of TIDs, which is expected as the model factors in the RCI scenarios too. Considering TID values to be the theoretical measure of expected performance, the sequence of CAs is the same for both methods~: MaIS-CA$_1$ $>$ MaIS-CA$_2$ $>$ BFS-CA$_1$ $>$ BFS-CA$_2$. Assuming that the TID estimates are a reliable measure of CA performance, the Throughput$_{Net}$ should decrease consistently with the rise in TID values, though not necessarily in a linear fashion. However, the two plots do not adhere to the expected pattern and exhibit a marked deviation which raises a valid concern about TID being a reliable theoretical metric to predict the performance of a CA.

To validate this argument, we try to ascertain if there exists a strong correlation between the TID estimate of a CA and its performance in a given WMN. For a more comprehensive evaluation, we consider two more CA schemes in addition to BFS-CA and MaIS-CA. The first is a centralized heuristic scheme \cite{23Cheng} which initially assigns a common channel to all nodes in the WMN. It then replaces the channels currently assigned to links so that the overall \textit{interference number} of the network decreases \emph{i.e.}, the TID of the WMN decreases. We denote it by \textit{CEN-CA}. The second CA scheme is a maximal clique based approach \cite{17Xutao}, where the authors assign channels to radios in such a way that a \textit{maximal clique} of the wireless links can be formed. Maximal clique creation would ensure that the number of non-conflicting links in a WMN is high. The CA scheme is represented as \textit{CLQ-CA}. Further, we generate C-MMCG and E-MMCG variants for both CEN-CA and CLQ-CA, which are 
denoted in the naming convention being followed \emph{i.e.}, CEN-
CA$_1$ \& CLQ-CA$_1$ are the C-MMCG CAs, while CEN-CA$_2$ \& CLQ-CA$_2$ are their E-MMCG counterparts. Thus, we have a total of 8 CA schemes and for each CA, we compute the TID and observe its performance in terms of Throughput$_{Net}$ and PLR values.

 \begin{figure}[htb!]
\center 
\includegraphics[width=8.5cm, height=6cm]{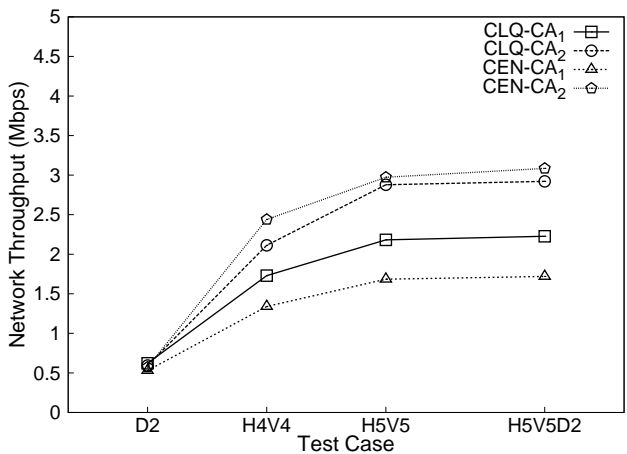}
\caption{Throughput$_{Net}$ of CEN-CA \& CLQ-CA in Test~Case~Class~3}
\label{G8}
\end{figure}
 \begin{figure}[htb!]
\center 
\includegraphics[width=8.5cm, height=6cm]{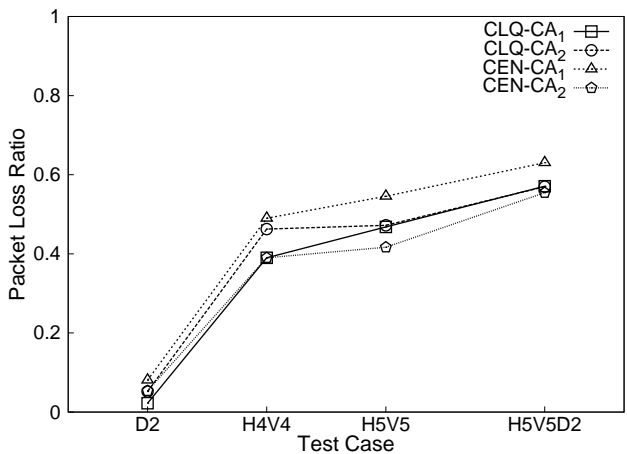}
\caption{PLR of CEN-CA \& CLQ-CA in Test Case Class 3}
\label{G8P}
\end{figure}
We subject the CA set to the test-cases of Test Case Class 3, as these are the test scenarios of peak network traffic and thus ideal for assessing impact of interference on the recorded metrics. For each CA, we record the Throughput$_{Net}$ and PLR values for the four test-cases \emph{viz.}, D2, H4V4, H5V5 \& H5V5D2. The results for BFS-CA and MaIS-CA have already been presented in the previous section. We now present the Throughput$_{Net}$ and PLR values registered for the new CAs in Figure~\ref{G8} and Figure~\ref{G8P}, respectively. Here too the E-MMCG CAs perform better than the corresponding C-MMCG CAs, both in terms of Throughput$_{Net}$ and PLR. The only exception is CLQ-CA$_2$ registering a marginally higher PLR than CLQ-CA$_1$ in test-cases D2 and H4V4. Nevertheless, the four E-MMCG CAs perform decidedly better than their C-MMCG counterparts and these results further consolidate the efficacy of E-MMCG in restraining the RCI.

 \begin{figure}[htb!]
\center 
\includegraphics[width=9cm, height=3cm]{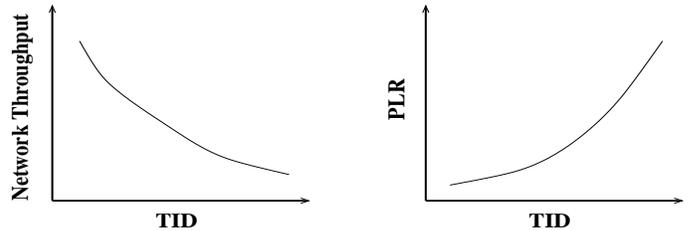}
\caption{Expected correlation of TID with Throughput$_{Net}$ and PLR}
\label{TID0}
\end{figure}

\begin{figure}[htb!]
\center 
\includegraphics[width=9cm, height=6cm]{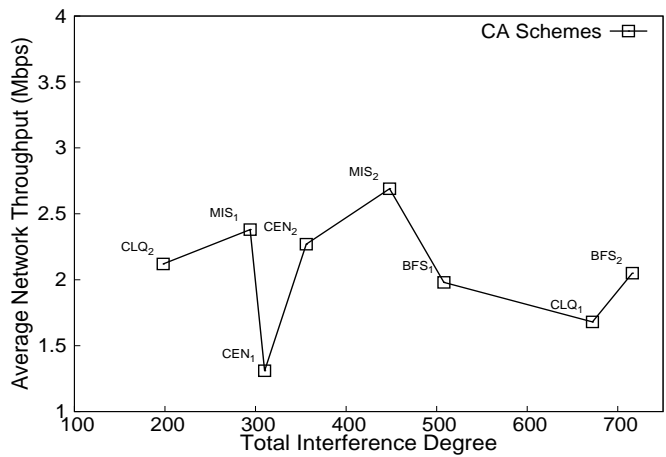}
\caption{Correlation of TID with Average Throughput$_{Net}$}
\label{TID1}
\end{figure}

\begin{figure}[htb!]
\center 
\includegraphics[width=9cm, height=6cm]{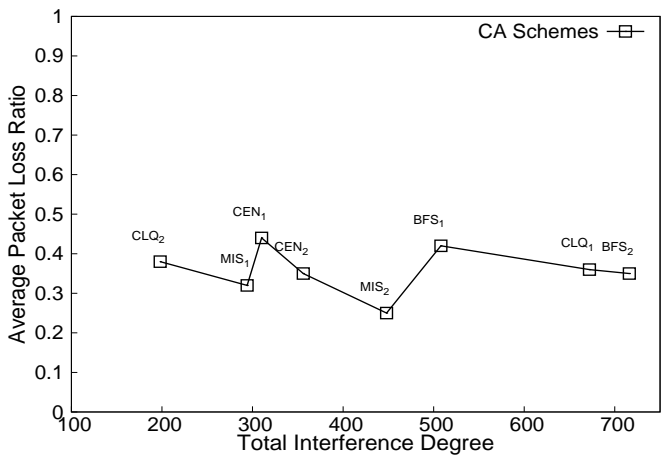}
\caption{Correlation of TID with Average Packet Loss Ratio}
\label{TID2}
\end{figure}

Next we compute the average of the recorded Throughput$_{Net}$ and PLR values of all the test-cases in Test Case Class 3, to determine the \textit{Average (Avg) Throughput$_{Net}$} and \textit{Average (Avg) PLR} for each CA. Average values of the two metrics reflect the overall performance of the CA in the grid WMN topology, which can be compared with the TID value of the corresponding CA. The TID values are computed using the E-MMCG algorithm. The computed values of Avg Throughput$_{Net}$ and Avg PLR are plotted against the TID values in Figs \ref{TID1} and \ref{TID2}, respectively. The CA schemes are denoted as: BFS-CA$_1$ (BFS$_1$), BFS-CA$_2$ (BFS$_{2}$), MaIS-CA$_1$ (MIS$_{1}$), MaIS-CA$_1$ (MIS$_{2}$), CEN-CA$_1$ (CEN$_{1}$), CEN-CA$_2$ (CEN$_{2}$), CLQ-CA$_1$ (CLQ$_{1}$) and CLQ-CA$_2$ (CLQ$_{2}$). 
Higher TID values suggest a greater adverse impact of interference. If we again assume TID estimates to be a reliable measure of CA performance, with increase in TID estimates the Avg Throughput$_{Net}$ values must decrease while the Avg PLR values must rise. As depicted in Figure~\ref{TID0}, the highest Avg Throughput$_{Net}$ value should correspond to the CA with the lowest TID estimate and thereafter the plot should exhibit a negative gradient \emph{i.e.}, a decrease in Avg Throughput$_{Net}$ as the TID value increases. Likewise, the Avg PLR values should increase as the TID estimate increases \emph{i.e.}, the plot should have a positive gradient. However, Figures \ref{TID1} \& \ref{TID2} show marked deviations from the assumed trends. This is evident from the fact that CLQ-CA$_2$ has the lowest TID estimate among the 8 CAs while it ranks 4\textsuperscript{th} in terms of Avg Throughput$_{Net}$ and has the 3\textsuperscript{rd} highest Avg PLR . In contrast, BFS-CA$_2$ has the highest TID value and yet 
offers a higher Avg Throughput$_{Net}$ and lower Avg PLR than 3 other CAs. We now present the variation in CA sequences when arranged in the increasing order of expected performance on the basis of theoretical TID estimates, and the increasing order of actual performance on the basis of observed metrics. The CA sequences are elucidated below.
\begin{itemize}
 \item \textbf{TID} : BFS-CA$_{2}<$ CLQ-CA$_{1}<$ BFS-CA$_{1}<$ MaIS-CA$_{2}<$ CEN-CA$_{2}<$ CEN-CA$_{1}<$ MaIS-CA$_{1}<$ CLQ-CA$_2$
 \item \textbf{Avg Throughput$_{Net}$} : CEN-CA$_{1}<$ CLQ-CA$_{1}<$ BFS-CA$_{1}<$ BFS-CA$_{2}<$ CLQ-CA$_2<$ CEN-CA$_{2}<$ MaIS-CA$_{1}<$ MaIS-CA$_{2}$
 \item \textbf{Avg PLR} : CEN-CA$_{1}<$ BFS-CA$_{1}<$ CLQ-CA$_2<$ CLQ-CA$_{1}<$ CEN-CA$_{2}<$ BFS-CA$_{2}<$ MaIS-CA$_{1}<$ MaIS-CA$_{2}$
\end{itemize}
It is clear that the CA performance sequence based on TID estimates is not in conformity with the CA performance sequences determined through experimental data. This finding will have profound implications on the numerous CA approaches proposed in research literature which assume a direct correlation between TID estimate and CA performance. The underlying idea in most of these interference-aware CA approaches is to minimize the local interference degree at a node, or minimize the TID while assigning channels to WMN radios \cite{Ding, Arunabha}. There is no doubt that the theoretical concept of \textit{Interference Degree} holds great relevance in estimating the intensity of prevalent interference at a node, or in an entire WMN. But extending this concept to compute \textit{TID} for a particular CA in a WMN, and predicting the expected behavior or performance of the CA on the basis of its estimated TID value is not a practically accurate approach. However, the task of ascertaining the benefit and relevance of 
predicating CA design on its TID estimate is beyond the scope of our current work.

\section{Conclusions And Inferences}
We begin by drawing the most fundamental conclusion that RCI has an adverse effect on the performance of WMNs. The \textit{Enhanced MMCG} model accounts for and adequately represents the RCI caused by spatially co-located radios. It is thus better equipped and algorithmically more tuned to alleviate the adverse impact of interference, than its conventional counterpart the \textit{Classical MMCG} model. The efficacy of a CA scheme in successfully mitigating the RCI depends on the extent of its knowledge of the conflict links that originate from RCI. The conventional MMCG generation techniques including the proposed C-MMCG approach, fail to account for and represent the RCI induced link conflicts in the MMCG model. The E-MMCG of a WMN remedies this problem by making the CA scheme aware of the prevalent RCI scenarios, thereby enabling it to assign channels to radios in a manner that restrains the RCI.\\
We can safely conclude that the CA deployments under the E-MMCG model invariably perform better than their peers under the C-MMCG model, for all the network performance indices. This is true for all 4 CA schemes we have considered in our study. The improvement noticed in MaIS-CA and CEN-CA is substantial as compared to BFS-CA and CLQ-CA where a less pronounced improvement was observed. Therefore, though the E-MMCG model augments the performance metrics of a CA, the underlying CA strategy also plays a determining role in this enhancement. This inference is a positive feature of the E-MMCG model that it does not alter or modify the inherent behavior or algorithmic nature of a CA scheme. 
On the use of \textit{TID} as a theoretical estimate of the prevalent interference in a WMN, we can infer that it is not an ideal metric, especially when used to predict the behavior of a deployed CA.
\section{Future Work}
Having established the notion of RCI and experimentally validated it, we plan to further explore the concept to engineer radio co-location aware channel assignment schemes. We also intend to take up the task of determining a better theoretical metric for estimating the impact of interference in a WMN and predicting CA performance.

\section{Compliance With Ethical Standards}
\subsection{Conflict of Interest} 
The authors declare that they have no conflict of interest. 

\subsection{Research involving human participants and/or animals}
This research work does not contain any studies with human participants or animals performed by any of the authors.

\subsection{Informed Consent}
No Individual Participants were a part of this study, nor was any personal data collected. Hence Informed Consent is not applicable to this study.
\bibliography{ref}

\end{document}